\documentclass[authoryear]{article}

\usepackage{amsmath}
\usepackage{amssymb}
\usepackage{graphicx}
\usepackage{natbib}
\usepackage{hyperref}
\usepackage{ifthen}

\usepackage{soul}

\usepackage{natbib}
\setcitestyle{abbrvnat,open={(},close={)}}

\usepackage[left=3cm,bottom=4cm,right=3cm,top=4cm]{geometry}

\def\affilKar{Institute for Hydromechanics, Karlsruhe Institute of Technology,
              Germany}

\def\affilTUW{Institute of Fluid Mechanics and Heat Transfer, TU Wien, 1060 Vienna, Austria }

\title{On the clustering of low-aspect-ratio oblate spheroids 
settling in ambient fluid}

\date{{\it Accepted in JFM 2023}}

\author{Manuel Moriche$^*$\footnotemark[2],
  Daniel Hettmann\footnotemark[2],
  Manuel Garc\'ia-Villalba\footnotemark[1]
 and Markus Uhlmann\footnotemark[2] \\
{\small
\footnotemark[2] \affilKar{}} \\
{\small
\footnotemark[1] \affilTUW{}} \\
{\small
corresponding author: {\tt manuel.moriche@tuwien.ac.at} }
}

\usepackage{booktabs}
\usepackage{pgfplots}
\pgfplotsset{compat=1.17}
\usepackage{myglossaries}
\glsdisablehyper

\newcommand{\revision}[2]{#2}
\newcommand{\revisions}[1]{}
\begin{document}

\maketitle
\begin{abstract}

We have performed particle-resolved direct numerical simulations
of many heavy non-spherical particles settling under gravity in the dilute regime.
The particles are oblate spheroids of aspect ratio $1.5$  and density ratio
$1.5$.
Two Galileo numbers are considered, namely $111$ and $152$, for which a
single oblate spheroid
follows a steady vertical and a steady oblique path, respectively.
In both cases, a strongly inhomogeneous spatial
distribution of the disperse phase in the form of columnar clusters is
observed, with a significantly enhanced average settling velocity as a
consequence.
Thus, in contrast to previous results for spheres,
the qualitative difference in the single particle regime does not result in
a qualitatively different behavior of the many-particle cases.
In addition, we have carried out an analysis of
pairwise interactions of particles in the well-known drafting-kissing-tumbling
setup, for oblate spheroids of aspect ratio
$1.5$ and for spheres.
We have varied systematically the relative initial position between the particle
pair and we have considered free-to-rotate particles and rotationally-locked ones.
We have found that the region of attraction for both particle shapes, with and
without rotation, is very similar. However, significant differences occur
during the drafting and tumbling phases.
In particular, free-to-rotate spheres present longer drafting phases
and separate quickly after the collision.
Spheroids remain close to each other for longer times after the collision,
and free-to-rotate ones experience two or more collision events.
Therefore, we have observed a shape-induced increase in the
interaction which might explain the increased tendency to cluster
of the many-particle cases.

\end{abstract}

keywords:
{\it
particle-laden flows, spheroids, clustering, collective effects, direct numerical 
simulation, immersed boundary method
}

\section{Introduction}
\label{sec:intro}
\begin{figure} 
\begin{center}
\includegraphics[scale=0.8]{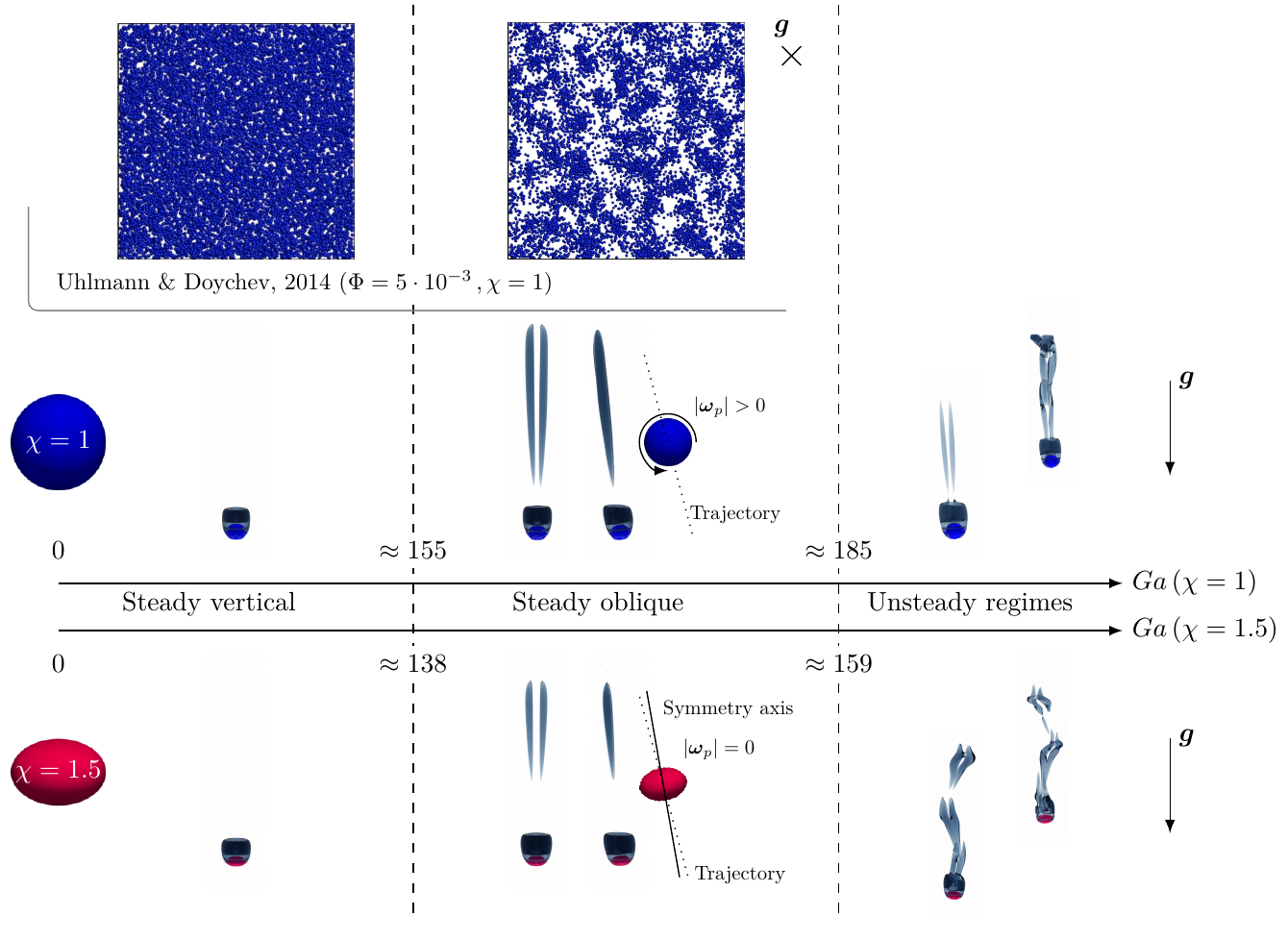} 
\end{center}
\caption{Single particle regimes for heavy ($\gls{kappa}=1.5$) spheres and
  oblate spheroids with aspect ratio $\gls{chi}=1.5$ as a function of
  the Galileo number $\gls{Ga}$.
  Reference data for dilute suspensions of spheres with the same
  density ratio and Galileo number from 
  \cite{uhlmann:2014a} are also included.
  The vortical flow structures in each regime are indicated with the
  aid of iso-surfaces of the \gls{f:Q}-criterion of \cite{hunt:1988}. 
  \label{fig:single_particle_regimes}} 
\end{figure}

Particle-laden flows
play an important role in natural and industrial systems, 
such as sediment transport in rivers, fluidized beds and
pollutants or hydrometeors in the atmospheric boundary layer.
It is well known that an isolated particle settling under gravity in
an otherwise ambient fluid exhibits a variety of regimes of motion
depending on its shape, mass density and size \citep{ern2012}.
Even for spherical particles diverse path regimes have
been observed, ranging from steady vertical to chaotic motion, and
including various intermediate states of different kinematic
complexity \citep[][cf.\ also
figure~\ref{fig:single_particle_regimes}]{jenny:2004,zhou:2015}.   
The transitions between the distinct regimes of particle motion are
the consequence of bifurcations in the flow pattern around the mobile  
particle, arising as the values of the governing parameters are varied. 
These parameters 
in the simplest single-particle case 
are the solid-to-fluid density ratio, $\gls{kappa}=%
\gls{rhop}/\gls{rhof}$, and the Galileo number, $\gls{Ga}=\gls{p:deq}\gls{Ug}/\gls{nu}$,
where \gls{p:deq} is the particle diameter (in the case of
non-spherical particles \gls{p:deq} is defined as the
diameter of a sphere with the same volume), 
$\gls{Ug}=\left(|\gls{kappa}-1|\left\|\gls{vg}\right\|\gls{p:deq}\right)^{1/2}$
is a gravitational velocity scale, \gls{nu} is the kinematic
viscosity, and \gls{vg} is the vector of gravitational acceleration.  
Whenever the volume fraction (\gls{svf}) of the particulate phase
becomes non-negligible, particle-particle interactions can lead to
significant collective effects. For dilute systems these 
interactions are predominantly of indirect type, acting through
long-range hydrodynamic forces. For denser systems direct contacts 
between two or more particles come into play more frequently, thereby
contributing to the exchange of momentum. In the present work we are
interested in dilute systems (with solid volume fractions below one
percent), which is why we will concentrate on this regime in the
following literature review. 
Depending on the values of the parameter triplet
$(\gls{svf},\gls{kappa},\gls{Ga})$, dilute suspensions
can exhibit spatial particle distributions with significant
non-homogeneity.
A non-homogeneously distributed particle phase in turn is often
accompanied by important macroscopic effects, such as an altered
mean settling velocity and a modification of the induced fluid flow
features.

\begin{figure} 
\makebox[\textwidth][c]{ 
\includegraphics[scale=1.0]{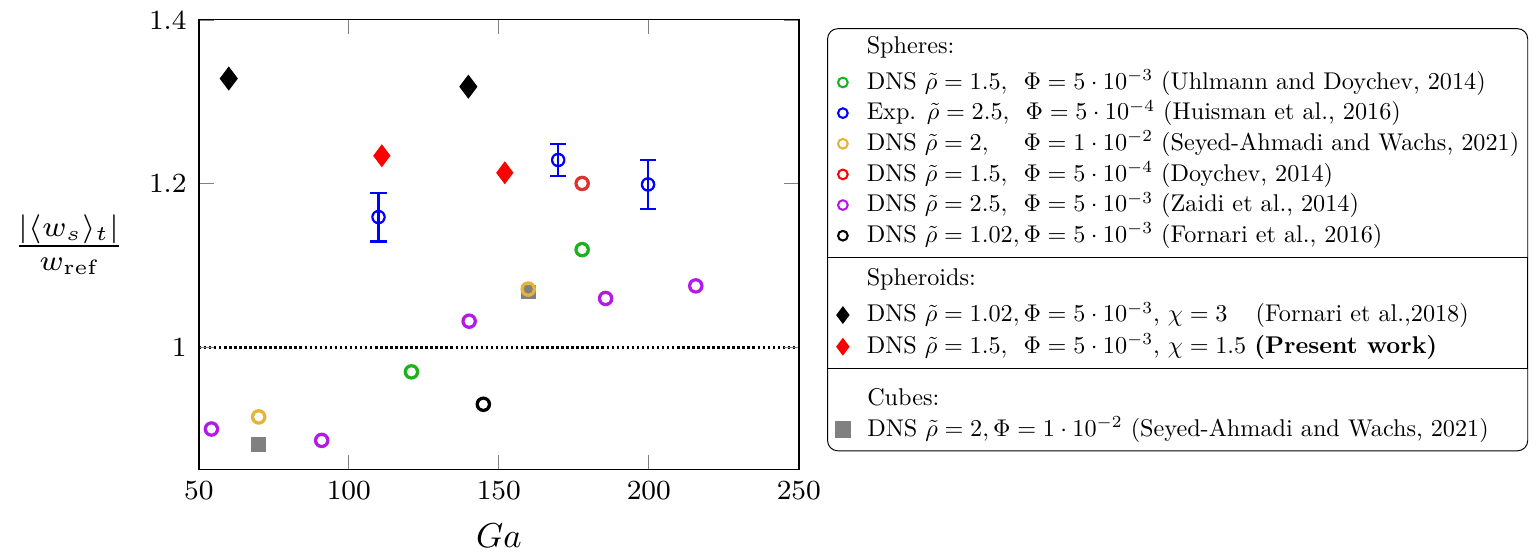} 
}
\caption{
Mean settling velocity versus Galileo number for spherical and non-spherical
suspensions of heavy particles with density ratio ${\cal O}(1)$ in the dilute
regime.
The velocity data is normalized with the corresponding mean settling
velocity of an isolated particle in the asymptotic (long-time) limit.
The error bars in the experimental data of \cite{huisman:2016} indicate
minimum and maximum values of the repetitions performed by the authors.
Present results are included for completeness.
\label{fig:settlingLiterature}} 
\end{figure}

At Galileo numbers of ${\cal O}(5\ldots 15)$ and density ratio $\gls{kappa}=2$
it has been
observed that dilute suspensions of spherical particles tend to form
horizontally-aligned pairs \citep{yin:2008}, while at larger 
Galileo numbers ($\gls{Ga}\approx 200$) and similar density ratios, where the
wake flow features a significant recirculation region, a vertical alignment is
more probable \citep{kajishima:2002,uhlmann:2014a}.  
\cite{kajishima:2002} were the first to investigate the formation of large, 
columnar-shaped particle clusters due to wake effects.
Their Particle-Resolved Direct Numerical Simulations (PR-DNS) in triply-periodic
boxes showed that clusters tend to form for particle Reynolds numbers exceeding
$200\ldots300$ ($Ga\approx153\ldots210$), leading to strongly enhanced average
settling velocities.
In their initial work, the particle rotation was suppressed for computational
simplicity, and a density ratio $\gls{kappa}=8.8$ was considered.
In the follow-up work of \cite{kajishima:2004}, the effect of particle rotation
was taken into account, and it was found that rotational motion leads to a lower
concentration of particles in clusters, since particles tend to escape a cluster
through a rotation-induced lift effect.
\cite{kajishima:2004} also determined a lower limit of the solid volume fraction
for the occurrence of clustering. 
The PR-DNS of \cite{uhlmann:2014a} were performed at a fixed density ratio 
$\gls{kappa}=1.5$, and it was observed that columnar clusters do form for 
$\gls{Ga}=178$ (which corresponds to an isolated particle in the steady oblique
regime), while no clusters are formed at $\gls{Ga}=121$ (steady vertical regime).
The authors suggest that the onset of clustering in dilute suspensions of
spherical particles is triggered by the bifurcation of the wake flow from steady
axi-symmetric to steady oblique, with the consequence that the particles in the
latter case drift horizontally (with random azimuthal angle) leading to an
enhanced probability of particle-particle encounters. 
Further data on collective effects upon clustering is available from the PR-DNS
studies of \cite{zaidi:2014}, \cite{fornari:2016} and \cite{seyed-ahmadi:2021},
which were all performed in similar triply periodic configurations, and from the
work of \cite{huisman:2016} who conducted experiments in a settling column with
glass beads in water. 
An overview of the average settling velocities from these various
data-sets for dilute suspensions of spherical particles in an
otherwise ambient fluid is given in
figure~\ref{fig:settlingLiterature}.
The PR-DNS results appear to give a consistent trend: for 
increasing values of the Galileo number the particles tend to settle 
at an average rate which is enhanced with respect to the velocity of
an isolated particle, and there appears indeed to be a cross-over
(from a reduced settling velocity to an enhancement) for
$Ga\approx150$.
The limited available experimental data in this parameter range is not
inconsistent with this picture, however showing a settling enhancement
already at lower Galileo number of $110$ and mild columnar cluster
formation.
\cite{huisman:2016} have identified the presence of the bounding 
container walls, which  presumably causes a large-scale recirculating
flow, as a possible cause for the observed discrepancy.

Dilute suspensions of non-spherical particles have received much less attention, mainly
due to the complexity associated with the particle shape 
and the corresponding increase in the size of the parametric space.
Spheroids have been the preferred choice of several authors, partly due to their
convenient parametrization.
Their shape is defined uniquely by the aspect ratio $\gls{chi}=\gls{dd}/%
\gls{aa}$, where \gls{dd} and \gls{aa} are the equatorial diameter and the
length of the symmetry axis, respectively (see figure
\ref{fig:problem_description}a,b).
The spheroidal shape allows for a smooth transition between flat,
disk-like shapes (for very large values of $\gls{chi}$), and elongated, fiber-like
geometries (for very small values of $\gls{chi}$), while 
a sphere is recovered for $\gls{chi}=1$.
When oblate spheroids ($\gls{chi}>1$) are considered, the regime maps
of a single particle are more complex when compared to spheres
\citep{zhou:2017}. 
Despite this increase in complexity, specific combinations of
\gls{chi} and \gls{kappa} result in a single oblate spheroid exhibiting the so-called 
``sphere-like scenario'' in which the first bifurcation for
increasing \gls{Ga} is regular, transitioning from a vertical to an oblique
regime \citep[see figure
\ref{fig:single_particle_regimes},][]{zhou:2017,moriche:2021}.
\cite{fornari:2018b} studied the settling of dilute suspensions of almost
neutrally buoyant ($\gls{kappa}=1.02$) oblate spheroids with a moderately-flat
shape $\gls{chi}=3$.
They considered both dilute and dense regimes at Galileo numbers at which a
single particle follows a steady vertical path ($\gls{Ga}=60$) or an
unsteady path which is vertical in the mean ($\gls{Ga}=140$).
For their most dilute cases they observed a large enhancement of the mean
settling velocity compared to the single particle case, with only 
small differences between the two Galileo numbers (cf.\
figure~\ref{fig:settlingLiterature}).  
\cite{fornari:2018b} have also detected non-homogeneous particle
distributions in the form of vertical particle trains with lateral
dimensions of the order of $10$ times the length of the symmetry axis,
based upon visualization and on the analysis of pairwise distribution
functions.  
The authors attributed the enhancement of settling velocity to the
occurrence of particle-pair interactions which lead to the formation
of piles of particles which practically stick to each other in the
case of oblate spheroids with $\gls{chi}=3$.

Turning now to other shapes, \cite{seyed-ahmadi:2021} more recently studied the
behavior of settling cubes with $\gls{kappa}=2$, $\gls{svf}=0.01\ldots0.2$, and
Galileo numbers for which a single cube tends to a steady oblique path with very
small tilting angle ($\gls{Ga}=70$) and to a helical path ($\gls{Ga}=160$) for
large times.
For their smallest solid volume fraction these authors report an increase in the
relative settling velocity, which is similar to what is observed in suspensions
of spheres (cf.\ figure~\ref{fig:settlingLiterature}).
Interestingly, \cite{seyed-ahmadi:2021} observe that the intensity of columnar
cluster formation (measured with the aid of particle-pair distribution functions)
is somewhat lower in the case of cubes as compared to the spherical counterpart.
This effect is attributed by the authors to higher rotation rates in the former
case, leading to an enhanced lift force that increases their probability to
escape from an existing cluster.
As noted above in the case of spheres, the settling regime of a single
particle appears to be relevant to the macroscopic behavior of dilute
suspensions. 
As a next step towards the understanding of collective effects, it is useful to
investigate the interaction of settling particle pairs, a setup which may lead to the 
so-called \gls{dkt} process \citep{fortes:1987}.
For sufficiently high Galileo number a trailing sphere 
initially released inside of the wake region of a leading sphere 
will approach the latter one during the drafting phase, they will
touch (`kissing'), and then interchange positions (`tumbling'), 
since a vertically-aligned pair of spheres is unstable, before
eventually separating.  
The reduced size of the problem compared to the many-particle cases makes 
\gls{dkt} simulations
a useful laboratory 
to understand the underlying physics of many-particle cases.
Indeed, \cite{fortes:1987} used auxiliary cases with the \gls{dkt} setup to
support their results on the spatial structure of many spherical
particles in a fluidized bed. 
Regarding non-spherical particles, \cite{ardekani:2016} observed a suppression
of the tumbling phase in pairwise interactions of moderately flat oblate
spheroids ($\gls{chi}=3$), as well as an increased collision domain.
The authors conjectured that the essentially infinite interaction time
of piled up particles would lead to enhanced clustering in the
corresponding many-particle case. 
This conjecture was later indeed confirmed by \cite{fornari:2018b}.
The numerical simulations of \gls{dkt} cases in the literature use a
common configuration: two particles settling in an initially quiescent fluid
inside a container, whose vertical size should be large enough to accommodate
the different stages of the \gls{dkt} case, and also minimize the (undesired)
influence of the lower boundary of the computational domain.
The vertical size of the container varies from $20\gls{p:deq}$
\citep{glowinski:2001,breugem:2012} to approximately $40\gls{p:deq}$
\citep[][]{patankar:2000,ardekani:2016}.
The main drawback of this configuration is that any modification of the
parameters (for example the relative initial position between the particles)
which increases the duration of the drafting phase, would require a further
increase of the computational domain.
As a result, the simplicity of the \gls{dkt} setup is somehow diminished.
Using periodic boundary conditions in the vertical direction, as some authors
have done for single particle cases \citep{kajishima:2002,doychev:2014}, also 
demands large domains in order to minimize wake effects from periodic repetitions.
To overcome these difficulties, in the present work we propose to use 
inflow/outflow boundary conditions along the vertical direction
along with a carefully adjusted vertical velocity at the inflow. 
Thus, the settling particles are not perturbed by their own wakes, and
the domain size becomes less of a critical parameter. 
To the best of our knowledge, this strategy has not been employed
previously.  
Here it allows us to employ moderately small computational domains, so that 
we can perform a campaign of simulations varying the relative initial position
of the particle pair and, therefore, we can obtain statistically significant results.
In the present work we analyze the clustering behavior of dilute suspensions of
spheroids with a relatively low aspect ratio $\gls{chi}=1.5$, a density ratio
$\gls{kappa}=1.5$ and two Galileo numbers $Ga=\{111, 152\}$. 
For these spheroids a single particle exhibits the so-called ``sphere-like
scenario'' \citep{zhou:2017}, and for suitable initial conditions a pair of
them undergo the three stages in a pairwise interaction (\gls{dkt}).
Our aim is to explore collective effects for non-spherical particle
shapes, 
while remaining relatively close to the well-explored spherical
shape.
The overarching question is: how do relatively small deviations
from the spherical shape affect the settling behavior of a dilute
suspension of rigid particles?
In addition to many-particle simulations in large domains, we perform
a large set of \gls{dkt} simulations of the mentioned spheroids as well as for the
reference case with spheres in which we analyze the effect of particle shape and of the 
angular motion by either suppressing or allowing the latter. 
The manuscript is structured as follows: in \S~\ref{sec:problem} we present 
the mathematical model describing the settling of particles in
unbounded, otherwise quiescent fluid and, in \S~\ref{sec:method} we
specify the numerical method and the physical and numerical
parameters. 
Results are presented in two steps: in \S~\ref{sec:results/multi} we focus on
many-particle cases, which represent the core of this  work, and in
\S~\ref{sec:results/dkt} we analyze a set of auxiliary simulations of a 
\gls{dkt} configuration in order to support the observations
made in the many-particle cases.
In \S~\ref{sec:discussion} we discuss the implications of the
\gls{dkt} results for the many-particle case. 
Finally, a summary and conclusions can be found in \S~\ref{sec:conclusions}.
\section{Problem description}
\label{sec:problem}
\begin{figure} 
\makebox[\textwidth][c]{ 
\includegraphics[scale=1.0]{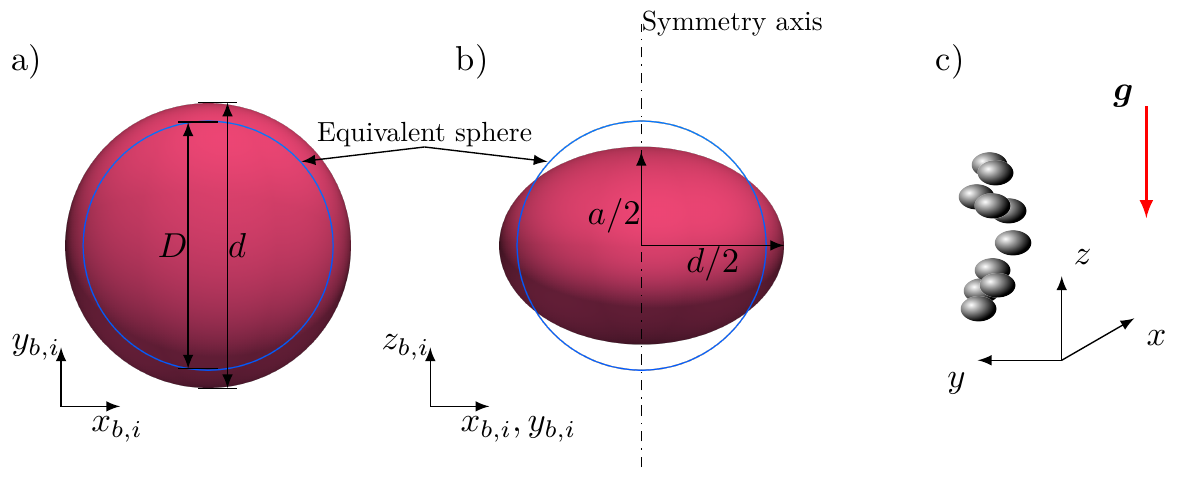} 
}
\caption{View of the $i$-th spheroid in its body-fixed reference system a) along
the symmetry axis and b) perpendicular to it. The blue line in a) and b)
represents a sphere with the same volume. c) Sketch of the problem in the global
reference system.
\label{fig:problem_description}} 
\end{figure}

We study the settling of particles under the action of gravity in an 
unbounded, initially quiescent fluid.
We consider an incompressible Newtonian fluid of density \gls{rhof} and 
kinematic viscosity \gls{nu}.
Particles are oblate spheroids of equatorial diameter \gls{dd} and aspect ratio 
$\gls{chi}=\gls{dd}/\gls{aa}$, where \gls{aa} is the length of their symmetry
axis (see figures \ref{fig:problem_description}a and b).
The particles are assumed to be rigid with homogeneous mass density \gls{rhop},
which is larger than that of the fluid ($\gls{rhop}>\gls{rhof}$).
The gravitational acceleration \gls{vg} is parallel to the vertical direction 
pointing in the negative $z$-direction, $\gls{vg} = -g\gls{ez}$, where
\gls{g} is the modulus of the gravitational acceleration (see figure 
\ref{fig:problem_description}c).

The fluid velocity $\gls{vu}=\left(\gls{vu_x},\gls{vu_y},\gls{vu_z}\right)$ is
governed by the Navier-Stokes equations for an incompressible, constant density
fluid 
\begin{subequations}\label{eq:gov}
\begin{align}
\ppt{\gls{vu}}+\left(\gls{vu}\cdot\nabla\right)\gls{vu}
              &= -\frac{\nabla \gls{p}}{\gls{rhof}} +\gls{nu} \nabla^2\gls{vu},
                \label{eq:gov_mom} \\
\nabla\cdot\gls{vu}&=0 \label{eq:gov_cont},
\end{align}
\end{subequations}
where \gls{p} is the pressure.
No slip and no penetration boundary conditions are imposed on the surface of each particle.
The linear and angular velocities of the particles,
$\gls{vup}=\left(\gls{vup_x},\gls{vup_y},\gls{vup_z}\right)$ and
$\gls{vomep}=\left(\gls{vomep_x},\gls{vomep_y},\gls{vomep_z}\right)$, 
respectively, are governed by the Newton-Euler equations 
\begin{subequations}\label{eq:motion}
\begin{align}
\gls{vp}\gls{rhop}\ddt{\gls{vup}}&=\int_{S} \gls{tau}\cdot\vec{n} \, \wrt \sigma
                            +\left(\gls{rhop}-\gls{rhof}\right)\gls{vp}\gls{vg} + \vec{F}, \label{eq:motion_lin}\\
\ddt{\left(\gls{I}\gls{vomep}\right)}&= %
 \int_{S} \gls{rs}\times \left(\gls{tau}\cdot\vec{n}\right) \wrt \sigma + \vec{T}, \label{eq:motion_rot}
\end{align}
\end{subequations}
where \gls{vp} and $S$ are the volume and the surface of the particle,
respectively, $\vec{n}$ is a unit vector normal to $S$ pointing towards the
fluid, \gls{tau} is the stress tensor ($\gls{tau}=-p\gls{eye}+\gls{rhof}\gls{nu}
\left(\nabla\gls{vu}+\nabla\gls{vu}^T\right)$), \gls{rs} is a position
vector with respect to the center of gravity of the particle and $\vec{F}$ 
and $\vec{T}$ are the solid-solid contact force and torque, respectively.

The problem is governed by four non-dimensional parameters, namely the
density ratio between the particles and the fluid, $\gls{kappa}=\gls{rhop}/%
\gls{rhof}$, the aspect ratio of the particles, \gls{chi}, the Galileo number,
\gls{Ga}, and the solid volume fraction, \gls{svf}.
The solid volume fraction is defined as $\gls{svf}=\gls{vol_part}/\left(
\gls{vol_part}+\gls{vol_fluid}\right)$ where \gls{vol_part} and \gls{vol_fluid}
represent the volume occupied by the particles and the fluid, respectively.
In this work we are interested in the dilute regime, and we set the solid volume
fraction to $\gls{svf}=5\cdot 10^{-3}$.
Regarding the shape of the particles, we select oblate spheroids of aspect ratio
$\gls{chi}=1.5$ which is a moderately small deviation from the spherical
reference geometry, that in turn has been extensively investigated in the past.
In particular, it is known that the settling regime map for oblate spheroids
with $\gls{chi}=1.5$ resembles that of a sphere \citep{moriche:2021}.
We fix the density ratio at $\gls{kappa}=1.5$ (which corresponds e.g.\ to some
plastic materials in water), and we select two values of the Galileo number: one
for which a single particle follows a steady vertical path ($\gls{Ga}=110.56$),
and another which leads to a single particle following a steady oblique path
($\gls{Ga}=152.02$).

\subsection{Definitions}
\label{sec:problem/definitions}

A velocity scale based on the gravitational acceleration, the density ratio and
the size of the particle can be defined as follows:
\begin{equation}\label{eq:problem/Ug}
\gls{Ug}=\sqrt{\left|\gls{kappa}-1\right|\gls{g}\gls{p:deq}}\,,
\end{equation}
where $\gls{p:deq}=\gls{dd}\,\gls{chi}^{-1/3}$ is the diameter of a sphere with
the same volume as the spheroid considered.
Based on the velocity scale \gls{Ug} \eqref{eq:problem/Ug} the Galileo number 
is defined as
\begin{equation}\label{eq:problem/Ga}
\gls{Ga}=\frac{\gls{Ug}\gls{p:deq}}{\gls{nu}} \,,
\end{equation}
and an a priori gravitational time scale can be formed as follows:
$\gls{tg}=\gls{p:deq}/\gls{Ug}$. 
For future reference let us define the relative velocity $\gls{vupri}=\left(
\gls{vupri_x},\gls{vupri_y},\gls{vupri_z}\right)$ of the $i$-th particle with
respect to the mean fluid velocity as:
\begin{equation}
\gls{vupri}(t)=\gls{vupi}(t)-\gls{vu_mf}(t),
\end{equation}
where the average operator $\langle \cdot \rangle_f$ indicates spatial
averaging over the entire domain occupied by the fluid $\Omega_f$, viz.
\begin{equation}\label{eq:avgFluid}
\langle \cdot \rangle_f =\frac{1}{\gls{vol_fluid}} \int_{\Omega_f} \left(\cdot \right)\wrt \gls{vx}.
\end{equation}
Let us also introduce the time-dependent settling velocity, averaged over the
set of particles as
\begin{equation}\label{eq:problem/defs/ws}
\gls{vup_ws}(t) = \gls{vupr_z_mp} (t) \,,
\end{equation}
where the average operator $\langle \cdot \rangle_p$ indicates the ensemble average
over the dispersed phase, which for a set of \gls{npart}
particles is expressed as  
\begin{equation}
\langle \cdot \rangle_p = \frac{ \sum_{i=1}^{\gls{npart}}%
    \left(\cdot\right)^{(i)}}{ \gls{npart} }\,.
\end{equation}
Similarly, the standard deviation of the vertical and horizontal components of the
linear and angular particle velocity are defined as
\begin{subequations} \label{eq:vomep}
\begin{align}
\gls{vup_wstd}(t) &=\gls{vupri_z_sp} \,,  \label{eq:vomep_a} \\
\gls{vup_ustd}(t) &=\frac{1}{2}\left(\gls{vupri_x_sp}+\gls{vupri_y_sp}\right) \,, \\
\gls{mp:vomep_z;std}(t) &=\gls{vomepi_z_sp} \,, \\
\gls{mp:vomep_l;std}(t) &=\frac{1}{2}\left(\gls{vomepi_x_sp}+\gls{vomepi_y_sp}\right)  \,,
\end{align}
\end{subequations}
where the prime symbol indicates that the quantity is the fluctuating part
of the variable, defined as
\begin{equation}\label{eq:problem/fluct}
\left(\cdot\right)' = \left(\cdot\right) - \langle \cdot \rangle.
\end{equation}
Finally, it should be mentioned that the definition of the Galileo number used
in this work is equivalent to that used in \citet{fornari:2016b} and
\citet{ardekani:2016} for spheroids, and to that of \cite{seyed-ahmadi:2021} for
cubes.
We have selected this definition over the one used in some works dealing
exclusively with spheroids \citep{zhou:2017,moriche:2021} in order to
recover the same value of \gls{Ga} when working with spheres ($\gls{chi}=1$).
For a multiparticle case with a given parameter pair ($\gls{Ga},\gls{kappa})$, 
independently of its solid volume fraction \gls{svf}, we define the reference
velocity \gls{wref} as 
\begin{equation}\label{eq:wref}
\gls{wref}\left(\gls{Ga},\gls{kappa}\right) = \left| \gls{vupr_z} \right| \,, 
\end{equation}
where the value of \gls{vupr_z} in \eqref{eq:wref} is taken from the corresponding
single particle case. 
Please note that for the cases considered here the single particle case regime
is steady, therefore no (temporal or statistical) averaging is needed to compute
\gls{wref} in \eqref{eq:wref}.
According to the above definitions of \gls{vup_ws} and \gls{wref}, we define the
average particle Reynolds number 
\begin{equation}\label{eq:Re}
\gls{p:Redeq}=\frac{\left|\gls{mp:vup_ws;avg}\right|\gls{p:deq}}{\gls{nu}} \,,
\end{equation}
and the single particle Reynolds number
\begin{equation}\label{eq:Re0}
\gls{p:Redeq;0}=\frac{\gls{wref}\gls{p:deq}}{\gls{nu}} \,,
\end{equation}
where the operator $\langle \cdot \rangle_t$ indicates temporal averaging.
\section{Methodology}
\label{sec:method}

\subsection{Numerical method}

The Navier-Stokes equations \eqref{eq:gov} are integrated in time
by means of a three-stage Runge-Kutta scheme in which the viscous term is 
treated implicitly and the advective term explicitly \citep{rai:1991,verzicco:1996}.
The fractional step method proposed by \cite{brown:2001} is used to fulfill the
continuity constraint.
Spatial derivatives are approximated with central finite-differences of 
second order on a staggered, uniform, Cartesian grid.

The presence of the body is modeled by the direct-forcing immersed boundary
method proposed by \cite{uhlmann:2005} and later extended to track the 
motion of non-spherical particles by \cite{moriche:2021}, where extensive 
validation of the method can be found.
Collisions are modelled with a repulsive short-range normal
force (with a range $\gls{dx}$) such as to avoid non-physical overlapping of
particles (details can be found in \S~\ref{sec:collision}).

In a triply periodic setup, gravity continuously accelerates the system.
Therefore, in order to allow for a steady state, we add a constant-in-space
source term to the vertical momentum equation whose volume integral is equal
 (and opposite in sign) to the net force exerted by the particles on the fluid
\citep{hoefler:2000}.

\subsection{Computational setup}

\begin{table} 
\caption{Parameters of the present cases and of those in \cite{uhlmann:2014a}
\label{tab:cases}} 
\makebox[\textwidth][c]{ 
\begingroup \footnotesize %
\begin {tabular}{l|ccc|ccc|c}%
\toprule Case&\gls {chi}&\gls {Ga}&\gls {kappa}&$[L_x \text {x} L_y \text {x} L_z ]/\gls {p:deq}^3$&$N$&\gls {svf}&Work\\%
\toprule \texttt{G111}&\ensuremath {1.5}&\ensuremath {110.56}&\ensuremath {1.5}&\pgfmathprintnumber [fixed, precision=1]{5.4946289e1} x \pgfmathprintnumber [fixed, precision=1]{5.4946289e1} x \pgfmathprintnumber [fixed, precision=1]{2.1978511e2}&\ensuremath {6{,}336}&\ensuremath {5\cdot 10^{-3}}&Present\\%
\texttt{G152}&\ensuremath {1.5}&\ensuremath {152.02}&\ensuremath {1.5}&\pgfmathprintnumber [fixed, precision=1]{5.4946289e1} x \pgfmathprintnumber [fixed, precision=1]{5.4946289e1} x \pgfmathprintnumber [fixed, precision=1]{2.1978511e2}&\ensuremath {6{,}336}&\ensuremath {5\cdot 10^{-3}}&\\\toprule %
\texttt{M121}&\ensuremath {1}&\ensuremath {121.24}&\ensuremath {1.5}&\pgfmathprintnumber [fixed, precision=1]{6.8e1} x \pgfmathprintnumber [fixed, precision=1]{6.8e1} x \pgfmathprintnumber [fixed, precision=1]{3.4100006e2}&\ensuremath {15{,}190}&\ensuremath {5\cdot 10^{-3}}&Reference \\%
\texttt{M178}&\ensuremath {1}&\ensuremath {178.46}&\ensuremath {1.5}&\pgfmathprintnumber [fixed, precision=1]{8.5e1} x \pgfmathprintnumber [fixed, precision=1]{8.5e1} x \pgfmathprintnumber [fixed, precision=1]{1.7100006e2}&\ensuremath {11{,}867}&\ensuremath {5\cdot 10^{-3}}&(Uhlmann \& Doychev, 2014)\\%
\end {tabular}%
\endgroup %
}
\end{table} 

The computational domain is a cuboid with triply-periodic boundary conditions,
whose size is the result of a compromise between computational resource
requirements and physical realism.
Based on preliminary tests we choose a domain size of approximately 
$55\gls{p:deq}$ in the lateral directions and approximately $220\gls{p:deq}$ in
the vertical direction.
Please note that a full decorrelation in the vertical direction is not warranted
once clustering sets in, and in the horizontal direction once the clusters grow
to a size comparable of the computational domain.
This lack of full decorrelation, which was also observed in the case of 
corresponding spheres \citep{doychev:2014}, is further discussed in appendix
\S~\ref{sec:autocorr}. 
We select a spatial resolution of $\gls{p:deq}/\gls{dx}\approx 21$ ($\gls{dd}/%
\gls{dx}=24$), which is supported by the work of \cite{moriche:2021}.
In their work the authors show that compared to a spectral/spectral-element
solution the error in the mean settling velocity of a single spheroid of aspect 
ratio $\gls{chi}=1.5$, density ratio $\gls{kappa}=2.14$ at $\gls{Ga}=152.02$ is
smaller than $2\%$.
In the present case this results in a grid of $[1152 \times 1152 \times 4608]$
points.
Table \ref{tab:cases} shows the parameters of the two simulations presented
in this work.

\subsection{Initialization}
\label{sec:ini}
\revision{To}{In order to} 
initialize the flow around the particles we
\revision{run}{simulate} 
an initial transient
during a time interval of
$66.67\gls{p:deq}/\gls{vu_z_mf}$ in which the particles are fixed.
The initial position of the particles follows a random uniform distribution.
Their orientation is randomly distributed with a maximum deviation of $\pm5^\circ$
tilting angle with respect to the vertical axis.
The objective of constraining the angular position is to obtain a slight 
perturbation of the angular position with respect to the stable position of 
settling spheroids at moderate \gls{Ga}.
The same initial distribution of particles is used in both cases presented in
table \ref{tab:cases}.

During the fixed-particle transient we impose the Reynolds number of the flow
relative to the particles, \gls{p:Redeq;0}, as obtained (as an output parameter)
in the simulations with an isolated mobile particle at the target value of the
respective Galileo number, cf. table \ref{tab:cases}.
\revision{}{This is realized by means of a constant vertical pressure
  gradient, similar to the body force which counteracts the
  acceleration of the system when particles are freely mobile
  \citep{uhlmann:2014a}.} 
As it will be shown later, the flow around the particles during the initial 
fixed-particle transient mostly resembles that of the analogous single-particle 
case at the given \gls{p:Redeq;0}, due to the low concentration of
particles in both cases.
It should be mentioned that during this initial transient we obtain good 
decorrelation of all flow velocity components in all spatial directions (see
\S~\ref{sec:autocorr}).
\revision{}{Once the particles are released, the mean fluid flow relative
  to the particles is maintained through the above mentioned body
  force. Therefore, if perfectly adjusted, a single particle would not
  exhibit any vertical motion in the
  computational reference system, while the vertical motion of the
  many-particle ensemble is entirely due to mutual interactions
  \citep{uhlmann:2014a}.}
\section{Results}
\label{sec:results}

In this section we present the results obtained for many-particle cases after
particles are released.
We also present additional simulations of settling particle pairs in order to
analyze their ``drafting-kissing-tumbling'' dynamics.
\subsection{Clustering phenomena in many-particle cases}
\label{sec:results/multi}

\begin{table} 
\caption{Single-particle regime and time-averaged results of the
present cases and those in the work of \cite{uhlmann:2014a}}
\label{tab:cases_results}
\makebox[\textwidth][c]{ 
\begingroup \footnotesize %
\begin {tabular}{l|cc|ccccc|c}%
\toprule Case&Regime single part.&\gls {p:Redeq;0}&\gls {p:Redeq}& {$\frac {\gls {mp:vup_ws;avg}}{\gls {wref}}$} &$\frac {\gls {mp:vup_wstd;avg}}{\gls {wref}}$&$\frac {\gls {mp:vup_ustd;avg}}{\gls {wref}}$&$\frac {\gls {mp:vvort_s;avg}}{\gls {mp:vvort_s;rnd}}$&\\%
\toprule G111&steady vertical&\ensuremath {105}&\ensuremath {130}&\ensuremath {-1.2343}&\ensuremath {0.3753}&\ensuremath {0.1715}&\ensuremath {0.8695}&Present\\%
G152&steady oblique&\ensuremath {158}&\ensuremath {192}&\ensuremath {-1.2135}&\ensuremath {0.3806}&\ensuremath {0.1784}&\ensuremath {0.9282}&\\\toprule %
M121&steady vertical&\ensuremath {141}&\ensuremath {142}&\ensuremath {-1.0035}&\ensuremath {0.0924}&\ensuremath {0.0596}&\ensuremath {0.3452}&Reference (Uhlmann \\%
M178&steady oblique&\ensuremath {234}&\ensuremath {263}&\ensuremath {-1.125}&\ensuremath {0.1853}&\ensuremath {0.1249}&\ensuremath {0.6193}& \& Doychev, 2014)\\%
\end {tabular}%
\endgroup %
}
\end{table} 

When particles are released, they start to interact with their neighbors by
repeated drafting-kissing-tumbling events.
The interaction between particles is intense in both cases \verb!G111! and 
\verb!G152!, and rather similar (see animations in the supplementary material).
In the following we present: i) the enhanced settling and quantification of 
clustering, ii) the angular motion of the particles, iii) the arrangement of 
particles' trajectories and iv) a visualization of the main features of the 
flow.

\subsubsection{Enhanced settling and quantification of clustering}
\label{sec:results/multi/clustering}
\begin{figure} 
\makebox[\textwidth][c]{ 
\includegraphics[scale=1.0]{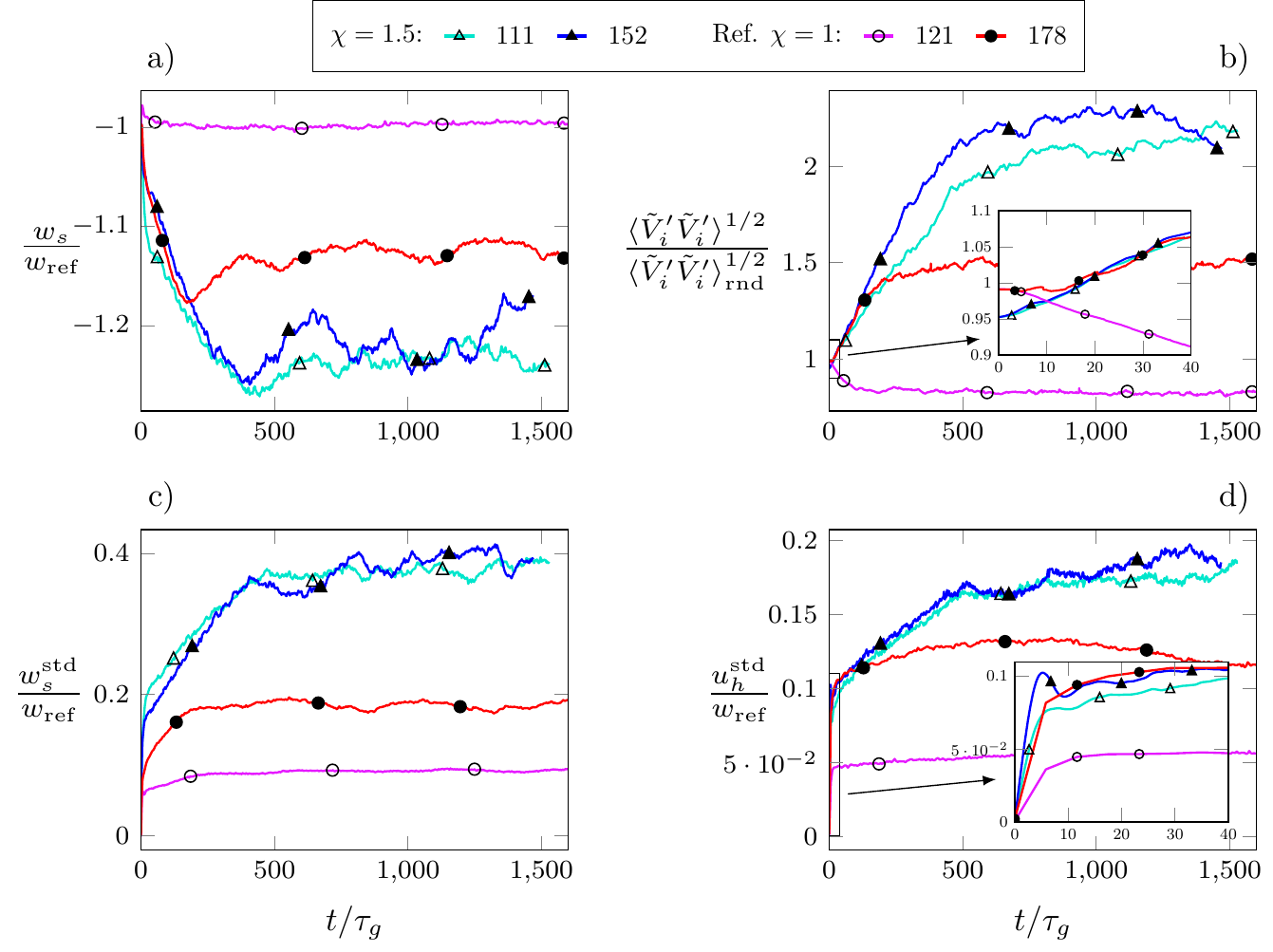} 
}
\caption{Time history of a) enhancement of the settling velocity (\gls{vup_ws}),
and standard deviation of c) settling and d) horizontal velocity, normalized
with the reference settling velocity from the single particle counterpart 
(\gls{wref}). b) Temporal evolution of the standard deviation of Vorono\"i cell
volumes (\gls{mp:vvort_s}), normalized with the value obtained for a random
Poisson process (\gls{mp:vvort_s;rnd}). Reference data for spheres is from 
\cite{uhlmann:2014a}.
\label{fig:wpvorstd}} 
\end{figure} 
Figure~\ref{fig:wpvorstd}a shows the time history of the average particle
settling velocity defined in \eqref{eq:vomep}.
It can be seen that in both present cases the ensemble of
mildly-oblate spheroids reaches on average a much enhanced settling
velocity magnitude after an initial transient of approximately
$400\gls{tg}$, after which the values saturate and continue fluctuating
around a value of roughly $-1.25$.
This behavior is quite in contrast to the known results for spheres (also
included in the figure for reference) for which a transition from expected
settling to enhanced settling occurs in the corresponding range of Galileo 
numbers, and for which the enhancement of the settling velocity was found to be
less pronounced (only roughly half of the present increase). 
Instead, the present suspension of spheroids with $\gls{chi}=1.5$ appears to
behave qualitatively similar to the much more flattened ($\gls{chi}=3$), almost
neutrally buoyant ($\gls{kappa}=1.02$) spheroids of \cite{fornari:2018b}, for
which the magnitude of the mean collective settling velocity was found to
increase by over $30$ percent. 

For completeness, let us report that the amplitude of the particle
velocity fluctuations (cf.\ figure~\ref{fig:wpvorstd}c,d) follows a
similar trend as the mean settling velocity, with a nearly linear
growth during the initial transient and subsequent fluctuations around
time-averaged values of approximately 
$\gls{vup_wstd} \approx 0.4\gls{wref}$, $\gls{vup_ustd} \approx
0.2\gls{wref}$. 
It should be noted that a clear signature of the oblique motion of
individual particles just after their release is visible in case
\verb!G152!, as can be seen in terms of an initial peak of the intensity of the
horizontal particle motion visible in the inset in figure~\ref{fig:wpvorstd}d.
After approximately $10\gls{tg}$ this peak disappears, and the two suspensions of spheroids at different
Galileo numbers exhibit very similar temporal evolutions. 

In order to investigate the spatial structure of the disperse phase,
we make use of the normalized Vorono\"i cell volume
\begin{equation}\label{eq:problem/defs/vvort}
\gls{mp:vvorti} =  \frac{\gls{mp:vvori}}{\gls{mp:vvor_mp}},
\end{equation}
where \gls{mp:vvori} is the volume of the $i$th cell of the  Vorono\"i
diagram obtained from a three-dimensional tessellation of space based
on the centroid locations of the particles
\citep{monchaux:2012,uhlmann:2014a}.  
In order to quantify the tendency to cluster we compare the standard
deviation of \gls{mp:vvorti} with the same quantity obtained in a
random Poisson process (RPP) of particles with the same shape and with
the same solid volume fraction.
For a mean concentration of particles such that $\gls{svf}=5\cdot 10^{-3}$,
these RPP reference values are $\gls{mp:vvort_s;rnd}=0.4176$ for
oblate spheroids of $\gls{chi}=1.5$ (determined in the framework of
the present work) and $\gls{mp:vvort_s;rnd}=0.4146$ for spheres
\citep[given in ][]{uhlmann:2020}. 
The graph in figure~\ref{fig:wpvorstd}b shows the temporal evolution
of the standard-deviation of the Vorono\"i cell volumes, normalized
with the reference value from RPP.
A direct qualitative correspondence with the temporal evolution of the
(magnitude of the) mean settling velocity in
figure~\ref{fig:wpvorstd}a can be observed. More specifically, after a
similar initial transient with an approximately linear growth of
\gls{mp:vvort_s}, the present suspensions of spheroids at both Galileo
number values reach very large clustering intensities, which by far
exceed those reported for spheres \citep[cf.\ case \texttt{M178}
from][]{uhlmann:2014a}. 
Based on the previous knowledge for settling spheres, and on the fact
that the present spheroids with $\gls{chi}=1.5$ are not too far from a
spherical shape, the strong clustering of the lower-Galileo case 
\verb!G111! is unexpected.

Next, we present time-averaged data of some of the time-dependent
quantities discussed above.
We discard the initial transient, and we start collecting statistics at
$t=500\gls{tg}$, resulting in a sampling time interval of approximately
$1000\gls{tg}$.
\begin{figure} 
\makebox[\textwidth][c]{ 
\includegraphics[scale=1.0]{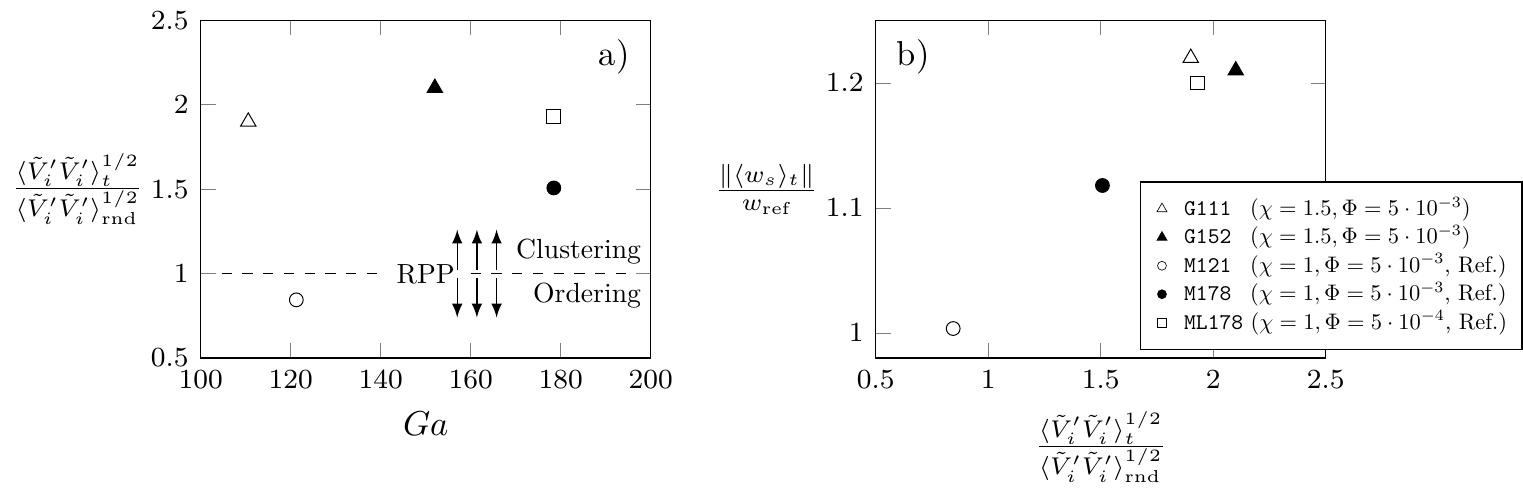} 
}
\caption{%
  (a) Time-averaged values of the standard deviation of Vorono\"i cell
  volumes, \gls{mp:vvort_s;avg}, normalized with the RPP reference
  values versus \gls{Ga}.
  (b) Magnitude of the time-averaged mean settling velocity,
  \gls{vup_ws}, versus \gls{mp:vvort_s;avg}. 
  Reference data for spheres from \cite{uhlmann:2014a} and
  \cite{doychev:2014}. 
  \label{fig:enhanced_settling_vs_clustering}
} 
\end{figure} 
Figure~\ref{fig:enhanced_settling_vs_clustering}a shows the
time-averaged values of \gls{mp:vvort_s} plotted versus the Galileo
number. It can be clearly seen that the present suspensions of
spheroids feature strong clustering with little effect of varying the
value of the Galileo number.
Interestingly, it turns out that the magnitude of the mean settling
velocity normalized by the single-particle reference value
is approximately proportional to the standard-deviation of
the Vorono\"i cell volumes normalized by its random value, i.e.\ to the
clustering intensity.
This relation is shown in
figure~\ref{fig:enhanced_settling_vs_clustering}b, for the present
spheroids and for the sphere suspensions of \citep{uhlmann:2014a} and of
\citep{doychev:2014}.
This observation should be checked with the help of additional data-sets in
the future, since -- if confirmed -- it might open up a possibility to
determine the mean settling velocity of a collective from knowledge on the
spatial structure of the dispersed phase alone. 

\begin{figure} 
\makebox[\textwidth][c]{ 
\includegraphics[scale=1.0]{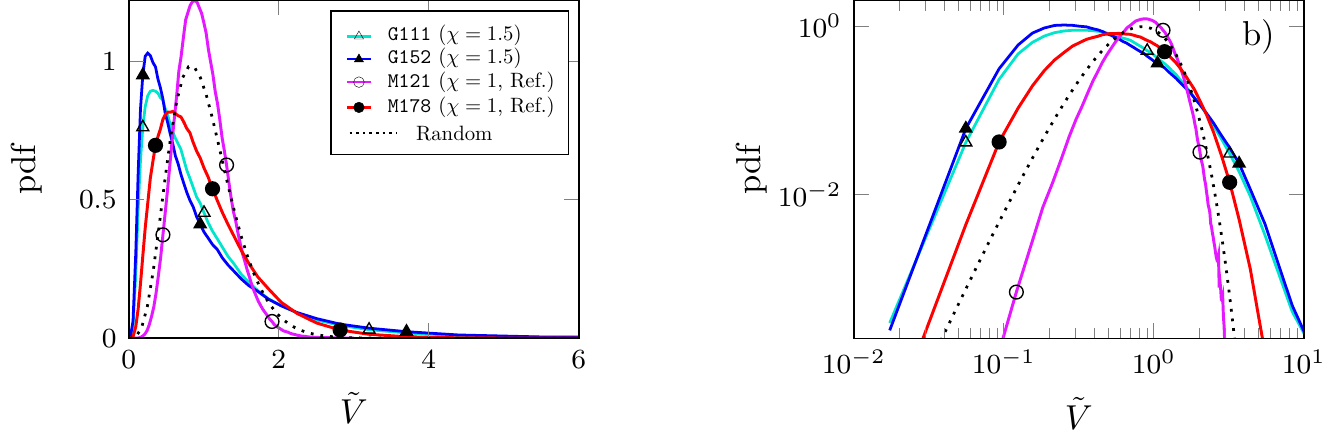} }
\caption{Probability density function of the normalized Vorono\"i cell volumes
in a) linear and b) logarithmic scale.
Reference data for spheres is from \cite{uhlmann:2014a}.
\label{fig:vorvol_Ga}} 
\end{figure} 
The distinct spatial arrangement of the spheroids compared to spheres can be further
characterized with the aid of the probability density functions (PDFs) of the
normalized Vorono\"i cell volumes \gls{mp:vvort}.
Figure \ref{fig:vorvol_Ga} shows the PDF of \gls{mp:vvort} of both cases with
spheroids (\verb!G111! and \verb!G152!) together with the PDF of a random
Poisson process \citep{uhlmann:2020} and the PDFs of two cases with spheres taken from 
\citet{uhlmann:2014a} (their cases \verb!M121! and \verb!M178!).
The same qualitative behavior is observed in both cases with spheroids: the
probability of finding volumes smaller than the average ($\gls{mp:vvort}=1$) is
high compared to a random Poisson process. 
This indicates that most of the particles form part of clusters.
Similarly, the probability of finding large volumes ($\gls{mp:vvort}\gtrsim 2$)
is higher than that of a random Poisson process, which implies the presence of
particles in large void regions.
Contrarily, case {\tt M121} from \cite{uhlmann:2014a} shows a tendency to a more
ordered state, where the probability of finding volumes similar to the average
value is higher.
\subsubsection{Angular velocity and orientation of particles}
\label{sec:results/multi/angular}

\begin{figure} 
\makebox[\textwidth][c]{ 
\includegraphics[scale=1.0]{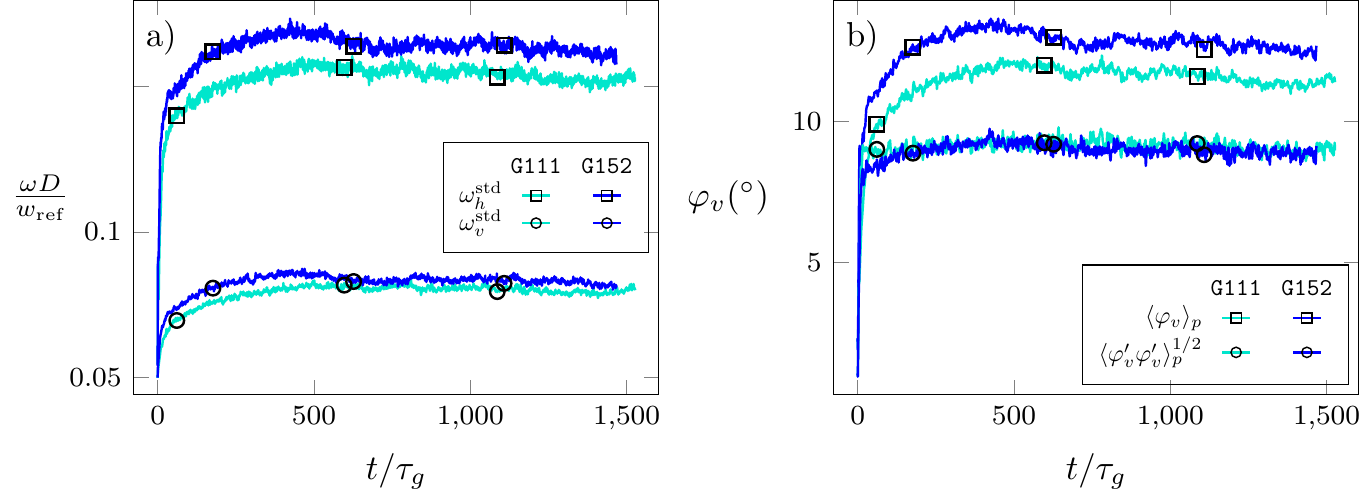} 
}
\caption{Time history of a) standard deviation of the angular velocity and b)
average and standard deviation of the orientation angle \gls{mp:az}.
\label{fig:angular_timehistory}} 
\end{figure}

Figure \ref{fig:angular_timehistory}a shows the time history of the standard 
deviations of the horizontal and vertical components of the angular velocity
of the particles
normalized with $\gls{wref}/\gls{p:deq}$.
Figure \ref{fig:angular_timehistory}b shows the time history of the averaged and
standard deviation of the tilting angle \gls{mp:az}, which is defined as the
angle between the symmetry axis of the spheroid and the vertical direction 
($0\le\gls{mp:az}\le 90^\circ$). 
The time evolution of these four quantities is remarkably different from
the time evolution of the quantities reported in Figure \ref{fig:wpvorstd}.
Within a few gravitational time units after the particle release all quantities
in figure \ref{fig:wpvorstd} increase from zero to values close to their 
asymptotic time-averages, after which they vary only very mildly.
This indicates that the angular motion of the particles shows less sensitivity
to collective effects than does the linear motion. 
Note that the converged values are slightly higher in case \verb!G152! when compared to 
case \verb!G111!, except for \gls{mp:az;std}, that presents values which are approximately equal in both cases.

\begin{figure} 
\makebox[\textwidth][c]{ 
\includegraphics[scale=1.0]{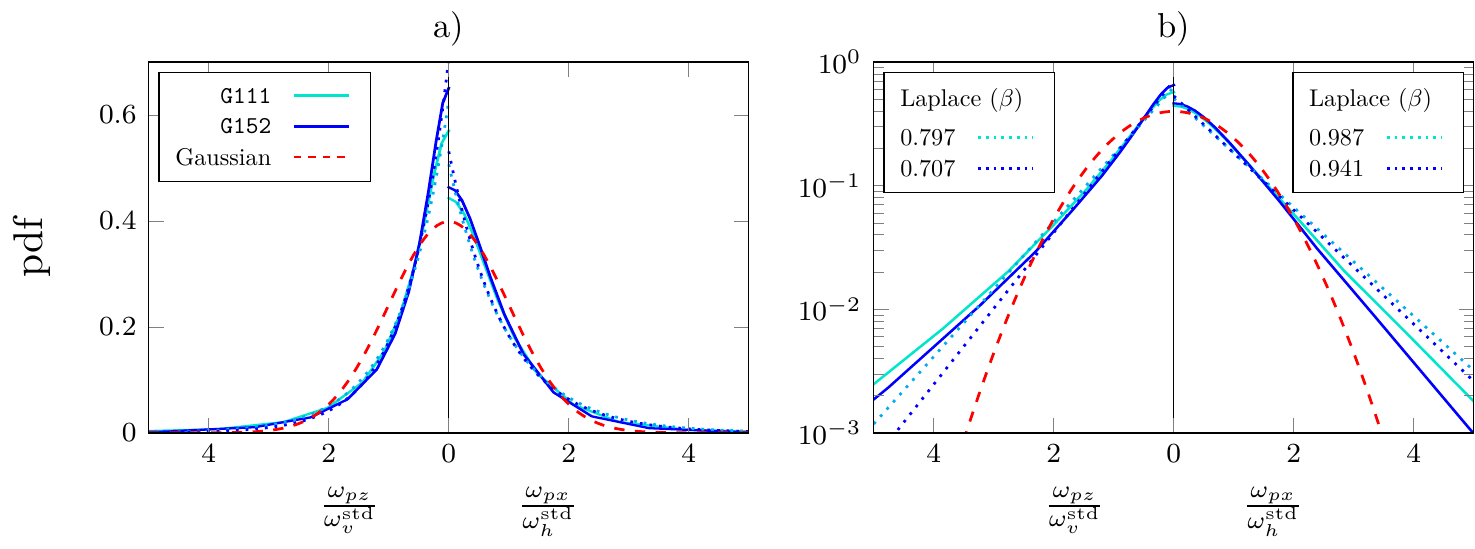} 
}
\caption{Probability density functions of the a) vertical and b) horizontal
components of the angular velocity.
The curves are fitted to a Laplace distribution whose parameter $\beta$ is 
indicated in the legend. The Gaussian curve is shown for comparison purposes.
\label{fig:angular_vel_pdf}} 
\end{figure}

\begin{figure} 
\makebox[\textwidth][c]{ 
\includegraphics[scale=1.0]{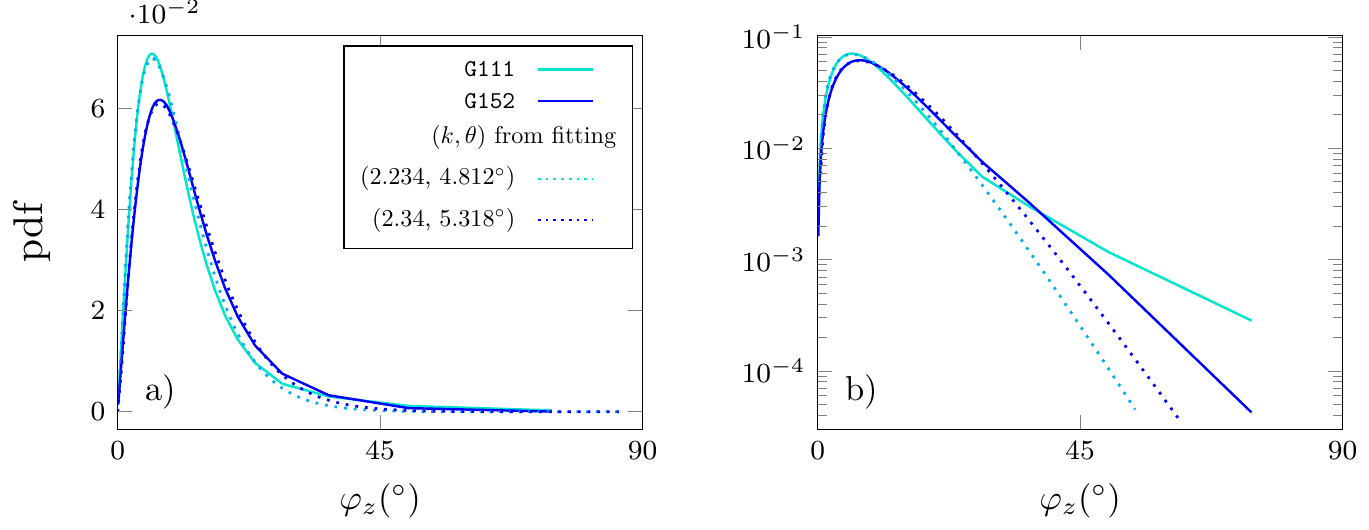} 
}
\caption{a) PDF of the tilting angle \gls{mp:az} of cases {\tt G111} and
{\tt G152} (the same information with the $y$ axis in logarithmic scale is shown
in b).
A fitted Gamma distribution is included (parameters from fitting included
in the legend). 
\label{fig:angular_az_pdf}} 
\end{figure}

Now let us focus on the probability density function of these angular
quantities.
Figure \ref{fig:angular_vel_pdf} shows the PDF of the vertical and horizontal
components of the angular velocity.
We have tried several known PDFs and we found Laplace's distribution
\begin{equation}
f\left(x;\beta\right) = \left(2\beta\right)^{-1}\exp\left(\left|x\right|/\beta\right)\,,
\end{equation}
with $\beta$ being a free parameter (cf. fitted numerical values in the figure),
the best fit to both components of the angular velocity. 
This highlights the exponential tails, i.e. the importance of extreme events of the
angular particle motion.
The vertical component shows, however, an approximately $20\%$ smaller value for
$\beta$ than the horizontal counterpart, indicating a more localized
distribution with higher kurtosis.
Similarly, we found the best fit of the estimated PDF of the tilting angle 
\gls{mp:az} with respect to the vertical to be a Gamma distribution
\begin{equation}
f\left(x;k,\theta\right) = %
 \frac{x^{k-1}}{\theta^k\,\Gamma(k)}\exp\left(-x/\theta\right) \,,
\end{equation}
where $k$ and $\theta$ are the shape and scale parameters.
Figure \ref{fig:angular_az_pdf}a shows the PDF of  \gls{mp:az} and the fitted 
Gamma distribution.
The obtained fitting shows very good agreement in the range
$0^\circ<\gls{mp:az}\lesssim30^\circ$.
For higher values $\gls{mp:az}>30^\circ$, the PDF of each case shows a lower
decay rate compared to the fitted Gamma distribution for both cases.
The shape parameter of the fitted Gamma distribution in both cases ($k\approx %
2.3$) indicates a fast, but non-abrupt approach to zero of the PDF in the limit
$\gls{mp:az}\rightarrow0^+$.
Please recall that the shape parameter of the Gamma distribution can infer the
following: when $k\leq 1$ the maximum probability is located at $\gls{mp:az}=0$
and then the PDF decreases monotonically as \gls{mp:az} increases, when $k>1$
the limit of the PDF as $\gls{mp:az}\rightarrow0^+$ is
zero (with strong gradients in the vicinity of $\gls{mp:az}=0$ for values of $k$
closer to unity), and when $k\geq3$ the distribution shows a slow increase of the
probability in the vicinity of zero (see figure \ref{fig:angular_az_pdf}b).
Finally, the non-intuitive zero value of the PDF in the limit
$\gls{mp:az}\rightarrow 0^+$ can be explained by the circumferential shape of
the bins used to generate it.
In order to obtain the PDF for a specific value of ${\gls{mp:az}}_0$, we define a 
bin by its edges $[{\gls{mp:az}}_a,{\gls{mp:az}}_b]$, where 
${\gls{mp:az}}_a < {\gls{mp:az}}_0 < {\gls{mp:az}}_b$.
In the limit $\gls{mp:az}\rightarrow 0^+$ and for increasing resolution of the
PDF (${\gls{mp:az}}_b - {\gls{mp:az}}_a\rightarrow 0$), the area on the surface
of a sphere between the edges tends to zero, and as a consequence the
probability of finding occurrences of $\gls{mp:az}$ such that
${\gls{mp:az}}_a < \gls{mp:az} < {\gls{mp:az}}_b$, also tends to zero.

\subsubsection{Trajectories}
\label{sec:results/multi/trajectories}

\begin{figure} 
\makebox[\textwidth][c]{ 
\includegraphics[scale=1]{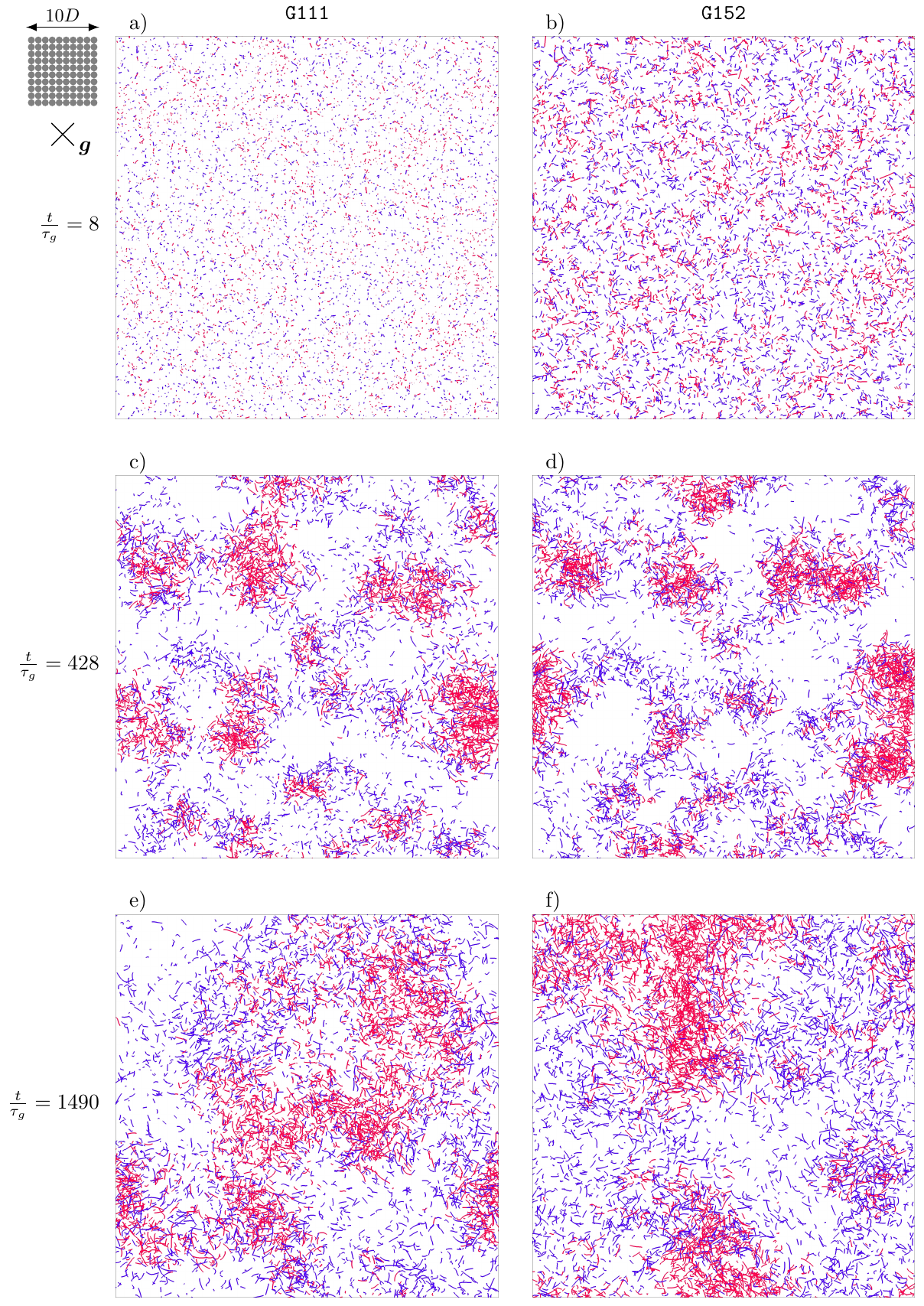} 
}
\caption{Top view of trajectories of the particles' center of gravity during a
time span of $[t_0-T_t,t_0]$, where $T_t=0.54\gls{tg}$ for cases {\tt G111}
(left column) {\tt G152} (right column). Trajectories are coloured according to
the particle's velocity relative to mean velocity of the mixture (red downwards,
blue upwards). 
\label{fig:vis_trajectories_top}} 
\end{figure}

\begin{figure} 
\makebox[\textwidth][c]{ 
\includegraphics[scale=1]{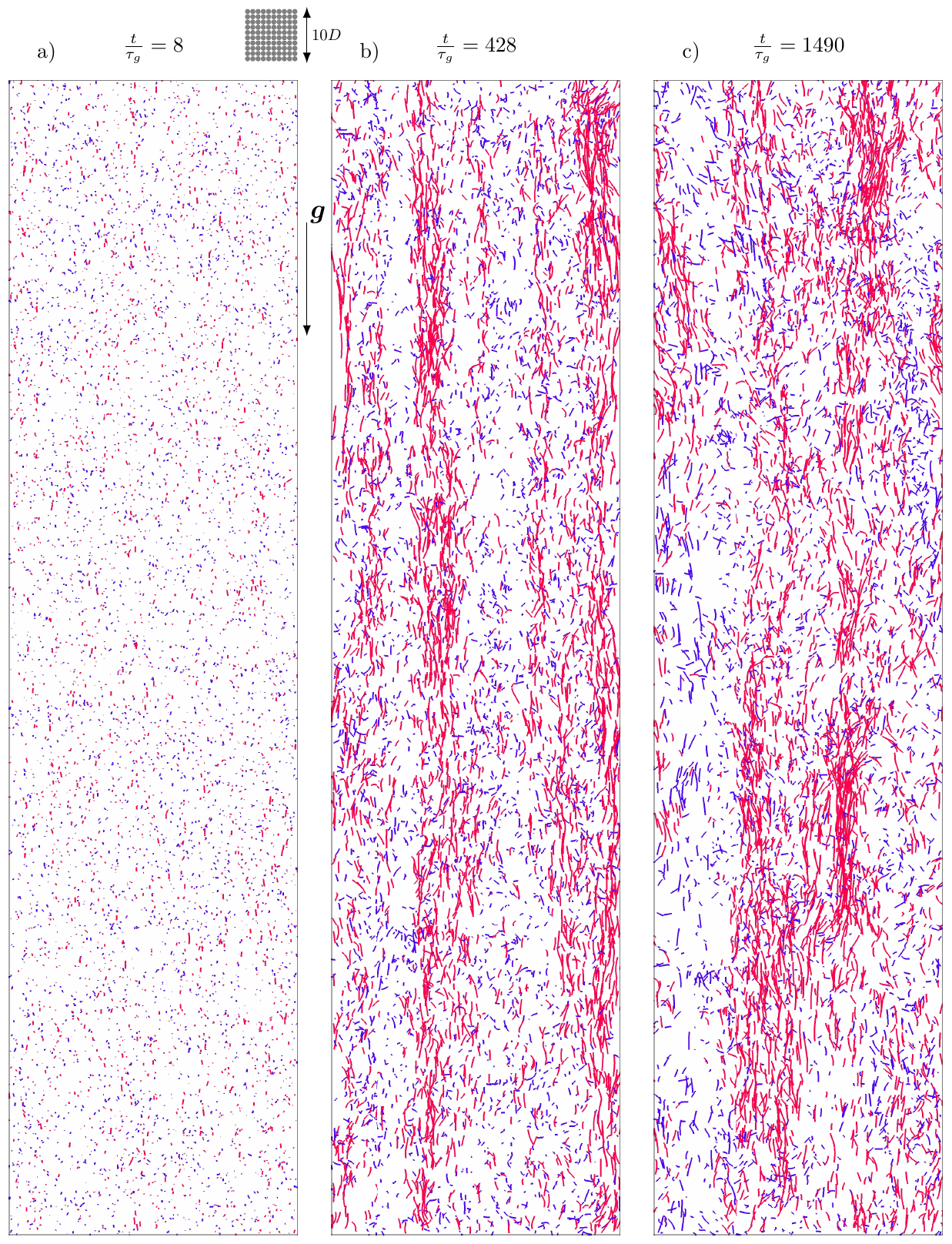} 
}
\caption{As in \ref{fig:vis_trajectories_top} but for case {\tt G111} only and
viewed from the side.
\label{fig:vis_trajectories_side_G111}} 
\end{figure}

\begin{figure} 
\makebox[\textwidth][c]{ 
\includegraphics[scale=1]{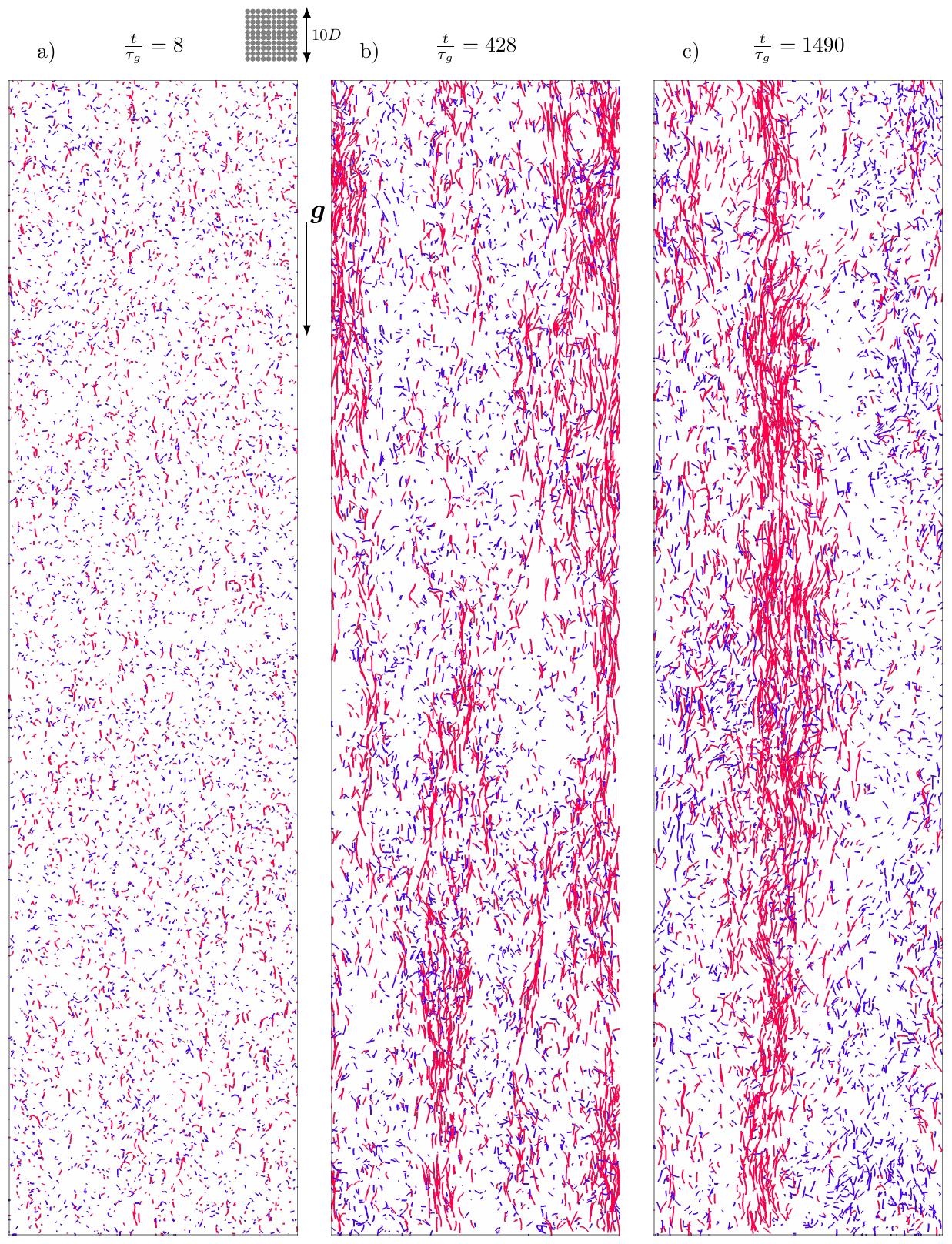} 
}
\caption{As in \ref{fig:vis_trajectories_top} but for case {\tt G152} only 
and viewed from the side.
\label{fig:vis_trajectories_side_G152}} 
\end{figure}

\def\mftop{\ref{fig:vis_trajectories_top}}
\def\mfsa{\ref{fig:vis_trajectories_side_G111}}
\def\mfsb{\ref{fig:vis_trajectories_side_G152}}

Now we turn our attention to the trajectories of the particles.
Figures \mftop{}, \mfsa{} and \mfsb{} show the trajectories followed by the 
center of gravity of the particles during a time span of $0.54\gls{tg}$ leading up to three
successive time instants in two perpendicular projections.
With this representation, considering the initialization procedure
(\S~\ref{sec:ini}) and if, hypothetically, collective effects were absent, 
particles would be represented by single points in the case \verb!G111! (single
particle follows steady vertical trajectory) and by
straight horizontal lines in the case \verb!G152! (single particle follows 
steady oblique trajectory).
The time instants selected are $t/\gls{tg}\approx 10, 400$ and $1500$, which 
correspond to: i) shortly after the particles are released, ii) the end of the
linear growth of \gls{vup_ws} and \gls{mp:vvort_s} (see figure
\ref{fig:wpvorstd}), and iii) the asymptotic state close to the end of the
simulated time, respectively.
Shortly after particles are released ($t/\gls{tg}\approx 10$), there is almost
negligible motion of the particles in case \verb!G111! (figures \mftop{}a and
\mfsa{}a show point-like trajectories) and a small lateral motion in case
\verb!G152! (figures \mftop{}b and \mfsb{}a show trajectories with a noticeable
horizontal component).
As both cases evolve, the trajectories show elongated shapes along the vertical
direction, indicating an enhancement of the settling velocity compared to the
single particle configuration.
On average, the length of the trajectories in the vertical direction keeps growing during the 
constant acceleration phase ($t\lesssim 400\gls{tg}$) forming columnar
clusters whose size is comparable to that of the computational domain
(figures \mftop{}c and d, \mfsa{}b and \mfsb{}b).
Again, the similarity of both cases is clearly noticeable.
There is, however, a tendency of the clusters in the case of lower Galileo to be
more stable compared to the case of higher Galileo when $t\approx 400\gls{tg}$
(see figures \mfsa{}b and \mfsb{}b).
We believe that the smaller flow disturbances of the low Galileo case allow the
presence of clusters with a cross section of the order of tens of \gls{p:deq} 
(see figure \mftop{}c) to fill the entire domain in the vertical direction,
whereas clusters of the same size in the horizontal direction of the higher
Galileo case (see figure \mftop{}d) are not stable and thus, appear to be more
localized.
This could explain the higher enhancement of the settling velocity with respect 
to the single particle case observed in the case at lower Galileo (figure 
\ref{fig:wpvorstd}a) but, it should be noted that this is only possible due to
the periodic configuration.
Therefore, the larger enhancement in \gls{vup_ws} observed in figure 
\ref{fig:wpvorstd}a for case \verb!G111! should be interpreted with care.
In the horizontal direction, clusters show a continuous growth until they reach
a size comparable to the computational domain (figures \mftop{}e and f).

\subsubsection{Flow visualization}
\label{sec:results/multi/visu}

\begin{figure} 
\makebox[\textwidth][c]{ 
\includegraphics[scale=0.75]{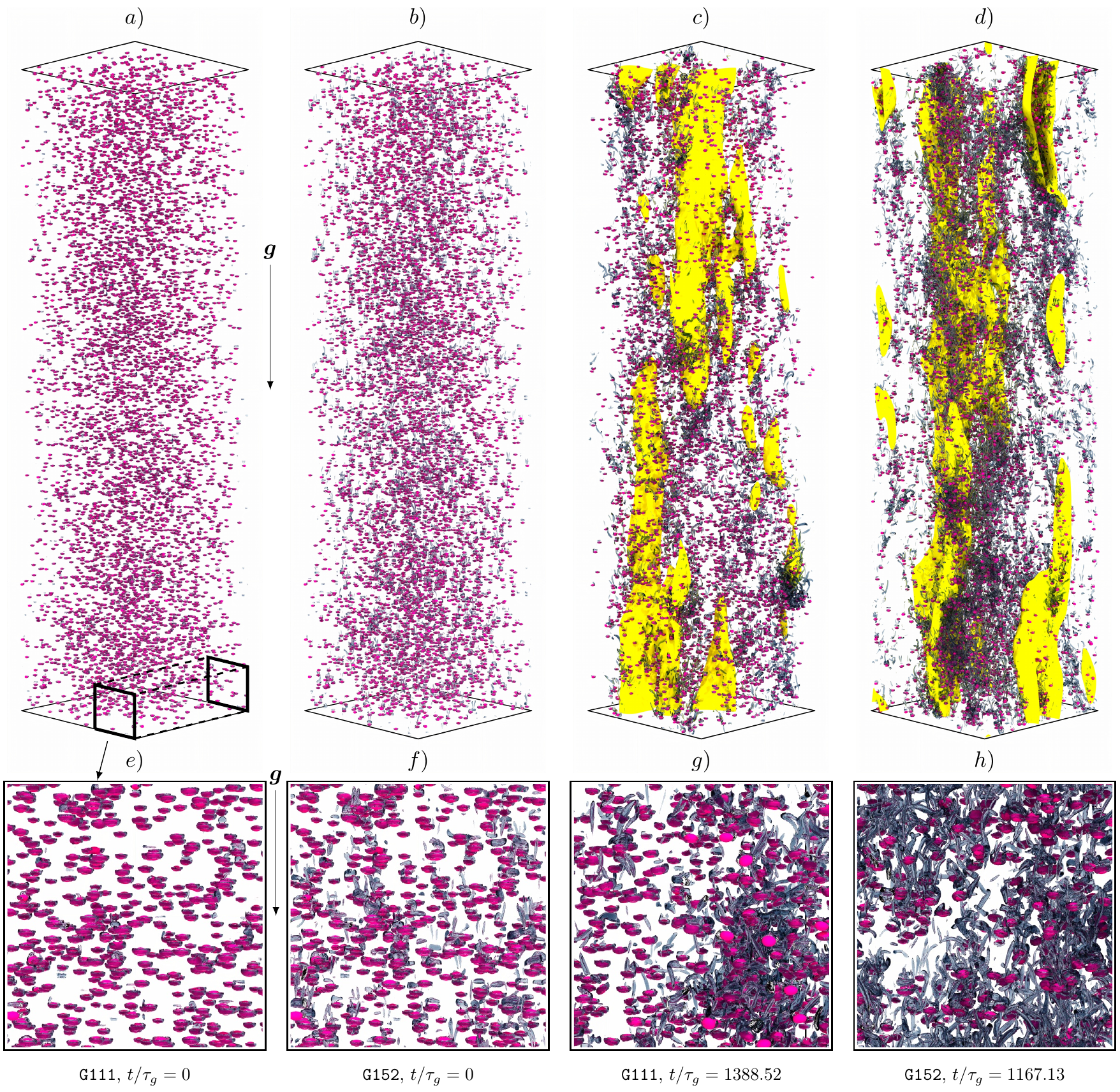} 
}
\caption{Visualization of isocontours of $\gls{f:Q}\gls{p:deq}^2/\gls{Ug}^2=0.7$
(case {\tt G111}) and $0.83$ (case {\tt G152}) and $\gls{f:vu_z;filt}=%
\gls{vu_z_mf}-0.5\gls{Ug}$.
Particles are represented in pink, isocontours of \gls{f:Q} with grey-colored 
surfaces and isocontours of \gls{f:vu_z;filt} with yellow surfaces.
Top row shows the whole domain (\gls{f:Q} and \gls{f:vu_z;filt}), 
bottom row shows a part of the domain (only \gls{f:Q}).
First and second columns correspond to the instant before the release of the
particles of cases {\tt G111} and {\tt G152}, respectively.
Third and fourth columns correspond to a converged state of each case.
\label{fig:visu}} 
\end{figure} 

Figure \ref{fig:visu} shows visualizations of the cases \verb!G111! and
\verb!G152! just before releasing the particles (panels a, b, e and f) and two
converged states (panels c, d, g and h).
In the figure the particles are represented together with isocontours of the 
second invariant (\gls{f:Q}) of the velocity gradient tensor \citep{jeong:1995}
and isocontours of the filtered vertical
velocity \gls{f:vu_z;filt}.
The filter applied to \gls{vu_z} is a Gaussian filter of width $2.3\gls{p:deq}$
used to visualize large-scale velocity fluctuations.
Just before particles are released ($t/\gls{tg}=0$) we observe that in
case \verb!G111! ($\gls{p:Redeq;0}=105$) the most common flow structure is a toroidal
vortex around each particle (figure \ref{fig:visu}e), whereas in case
\verb!G152! ($\gls{p:Redeq;0}=158$) we also frequently observe a double-threaded
wake (figure \ref{fig:visu}f).
The similarity of these vortical structures with the analogous single-particle
case at the given \gls{p:Redeq;0} is due to the low concentration of particles
in both cases.
After convergence has been reached ($t/\gls{tg}>1000$) both cases show large
regions of high-speed downward flow, which are absent when particles are 
released ($t=0$).
The strong clustering of both cases discussed above is clearly seen when
comparing the initial and converged snapshots of both cases. 
\subsection{Drafting-kissing-tumbling}
\label{sec:results/dkt}
\begin{figure} 
\begin{center}
\includegraphics[scale=0.8]{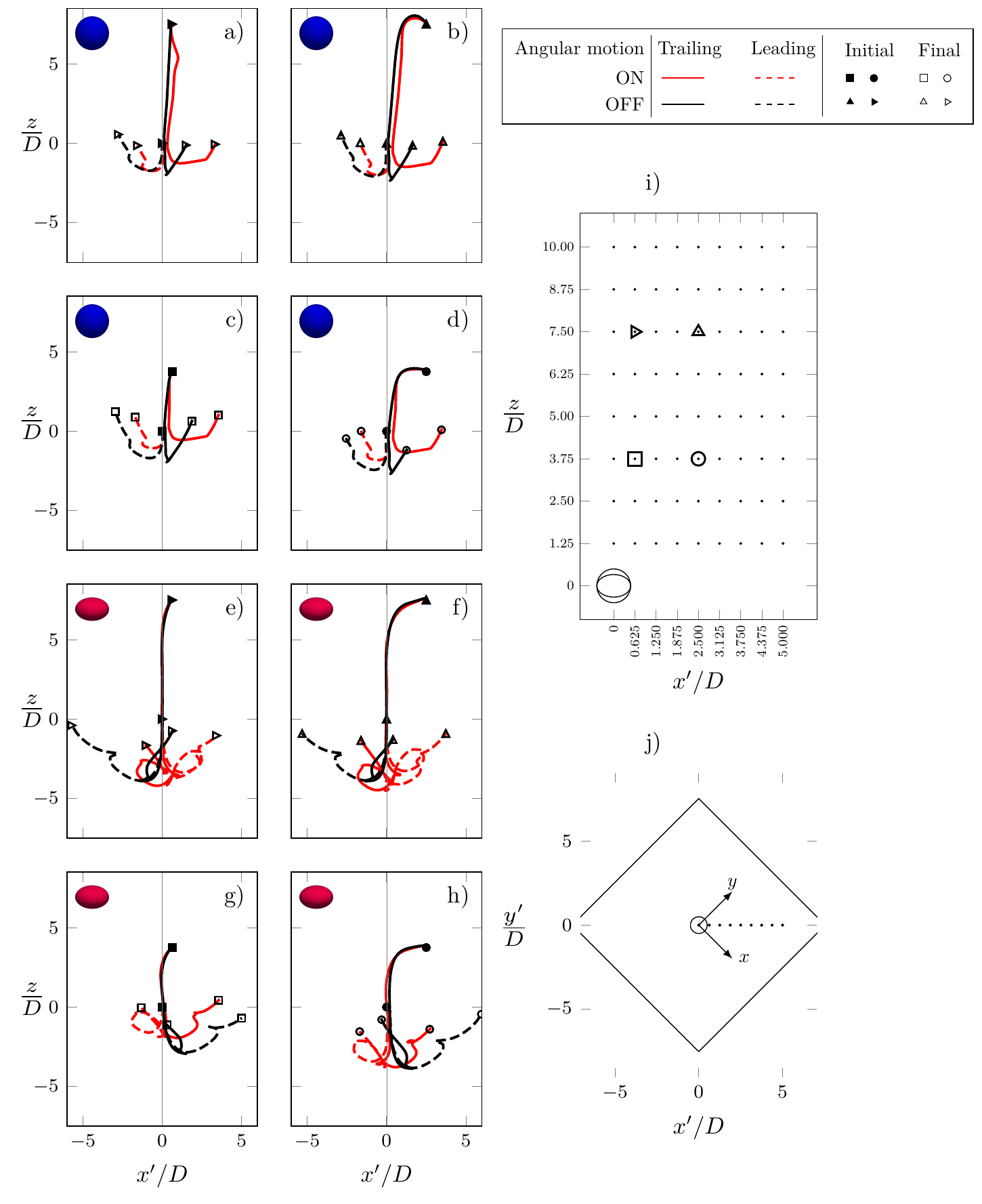} 
\end{center}
\caption{Trajectories of the trailing and leading particles for selected initial 
positions of spheres (a-d) and spheroids with $\gls{chi}=1.5$ (e-h), all with 
$\gls{kappa}=1.5$ at $\gls{Ga}=111$.
The reference frame is translating downwards at a constant speed slightly smaller than
the settling velocity of a single particle ($0.975\gls{wref}$).
Each panel contains the data of the cases with angular motion enabled and 
suppressed for a single initial condition and particle shape (see legend).
i) A sketch of the problem and the coordinates used is presented, in which the 
leading particle is represented with its actual shape and the trailing particle
with a marker.
The point markers correspond to all the initial conditions which we have computed,
and the symbol markers to those initial conditions which are shown in panels a-h.
j) Sketch of the $x',y'$ coordinates.
\label{fig:DKT_trajectories}} 
\end{figure}

Since animations show that interactions between particle pairs occur frequently
in the many-particle cases, we now proceed to an analysis of the interaction of
such pairs in isolation.
As we will see, these interactions are quite sensitive to the particle shape, 
and we believe that pairwise interactions are the key to understand the tendency
of oblate spheroids to cluster at Galileo numbers for which spheres do not
exhibit clustering.
We restrict the analysis to one combination of Galileo number and density ratio  
($\gls{Ga}=111$, $\gls{kappa}=1.5$) for which a single spheroid of aspect 
ratio $\gls{chi}=1.5$ and a single sphere result in a steady vertical regime 
\citep{jenny:2004,moriche:2021}.
As mentioned above, many-particle cases in the dilute regime ($\gls{svf}=%
5\cdot 10^{-3}$) with this combination of \gls{Ga} and \gls{kappa} present
strong clustering in the case of oblate spheroids with $\gls{chi}=1.5$ and 
no clustering in the case of spheres \citep{uhlmann:2014a}.
We include an additional series in this set of \gls{dkt} cases, in which the
angular motion is suppressed. Therefore, we have these four configurations:
\begin{itemize}
   \item Free-to-rotate spheres (angular motion enabled).
   \item Rotationally-locked spheres (angular motion suppressed).
   \item Free-to-rotate spheroids (angular motion enabled).
   \item Rotationally-locked spheroids (angular motion suppressed).
\end{itemize}
This is motivated by the study of \cite{kajishima:2004} who demonstrated that
suppressing the rotation of spheres leads to an enhanced clustering tendency.
We perform a parametric sweep of $72$ initial relative particle positions for
each configuration, resulting in a total number of $288$ simulations (see figure
\ref{fig:DKT_trajectories}i).
The problem setup is analogous to the setup of the many-particle cases presented
in \S~\ref{sec:method}, except that only two particles are present, and that we
use an inflow/outflow configuration in the vertical direction as described in
detail in appendix \ref{sec:dkt}.

Figure \ref{fig:DKT_trajectories} shows the trajectories of $16$ selected cases
projected onto a vertical plane intersecting the particles at their initial 
position. 
These cases correspond to different initial conditions of the four selected
configurations.
Please note that in these four cases particles do collide at least
once, while this is not the case for all initial separations
(as will be shown in figure \ref{fig:DKT_maps} below).
If we focus on the trajectories of the trailing particles, there is a clear
similarity in free-to-rotate and rotationally-locked spheroids: for these two
configurations the trailing particle of the four initial conditions shown in the
figure moves laterally until its center is vertically aligned with the leading
particle. 
Then, the trailing particle drifts towards the leading particle following an 
almost vertical path.
On the contrary, spheres present a different path along which the trailing
particle approaches the leading particle.
The trailing particle of both free-to-rotate and rotationally-locked spheres
presented in the figure drifts towards the leading one following 
an oblique path for a significant time during the final approach.

Interestingly, free-to-rotate spheres present a lateral shift (by roughly 
$\gls{p:deq}/4$) in their trajectory compared to their rotationally-locked
counterparts.
This is clearly evident in figure \ref{fig:DKT_trajectories}a, where the
trailing particle of the free-to-rotate pair first moves away laterally from the
leading particle, then making its way in a straight, oblique path towards it. 
\begin{figure} 
\begin{center}
\includegraphics[scale=0.9]{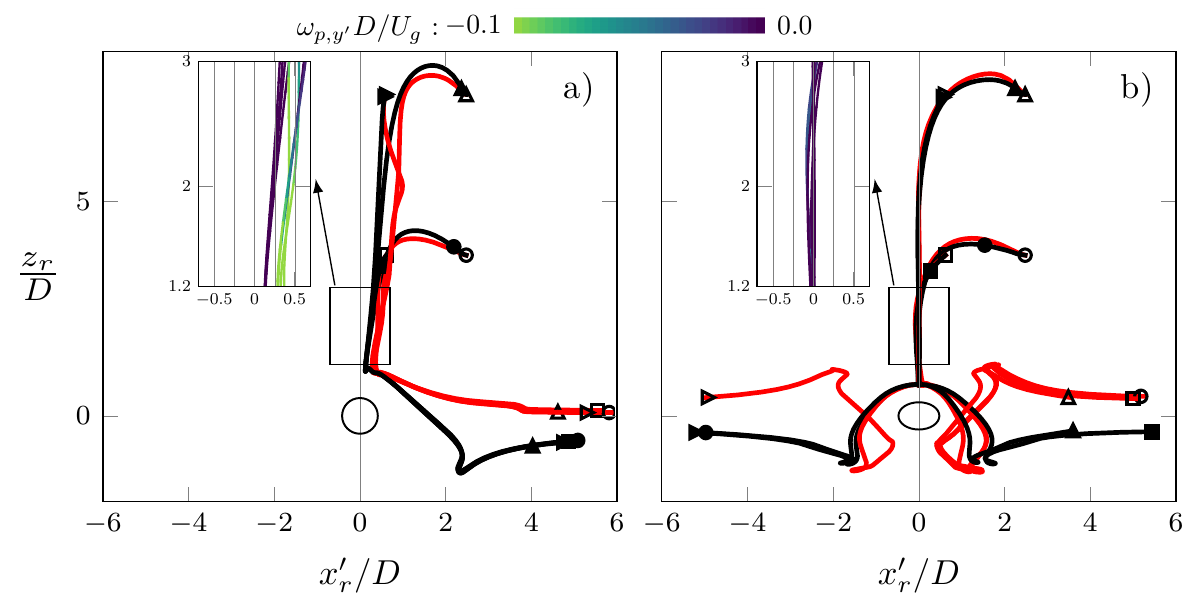} 
\end{center}
\caption{Trajectory of the trailing particle relative to the leading one for a)
spheres and b) spheroids. The close-up trajectories shown in the insets in a,b
are coloured with the angular velocity perpendicular to the plane shown (see 
legend). Line color and marker type follow the same convention as in figure 
\ref{fig:DKT_trajectories}.
\label{fig:DKT_trajectories_rel}} 
\end{figure} 
In figure \ref{fig:DKT_trajectories_rel} we show the trajectories of the 
trailing particle relative to the center of the leading particle for the same
configurations as presented in figure \ref{fig:DKT_trajectories}.
It can be seen how the rapid alignment along the vertical axis exhibited by the
spheroidal particles results in the possibility of finding the trailing particle
on either side of the vertical axis after the initial contact.
Spheres, on the other hand, which do not fully align vertically, do not cross
the vertical axis through the leading particle during the entire interaction.
A very interesting result is the robustness of the lateral shift of the trailing
particle for free-to-rotate spheres.
Independently of the four initial conditions shown in the figure,
the path followed by the trailing particle in the free-to-rotate sphere configuration
converges to a single master curve. 
This feature has been observed for a number of other initial conditions.
We attribute this lateral shift to a rotation-induced lift force resultant from
the finite angular velocity
(around the horizontal axis perpendicular to the plane shown in the figure)
reached by these particles (see inset of figure
\ref{fig:DKT_trajectories_rel}a).
The origin of the rotation of the particles ($\omega_{p,y'}<0$) is the shear
seen by the trailing particle because of the deficit in vertical velocity in the
wake of the leading particle.
A detailed analysis of the forces acting on a particle in motion in the wake of
another particle, however, is outside the scope of the present contribution, and
it is left as a worthwhile topic for future studies.
\begin{figure} 
\makebox[\textwidth][c]{ 
\includegraphics[scale=1.0]{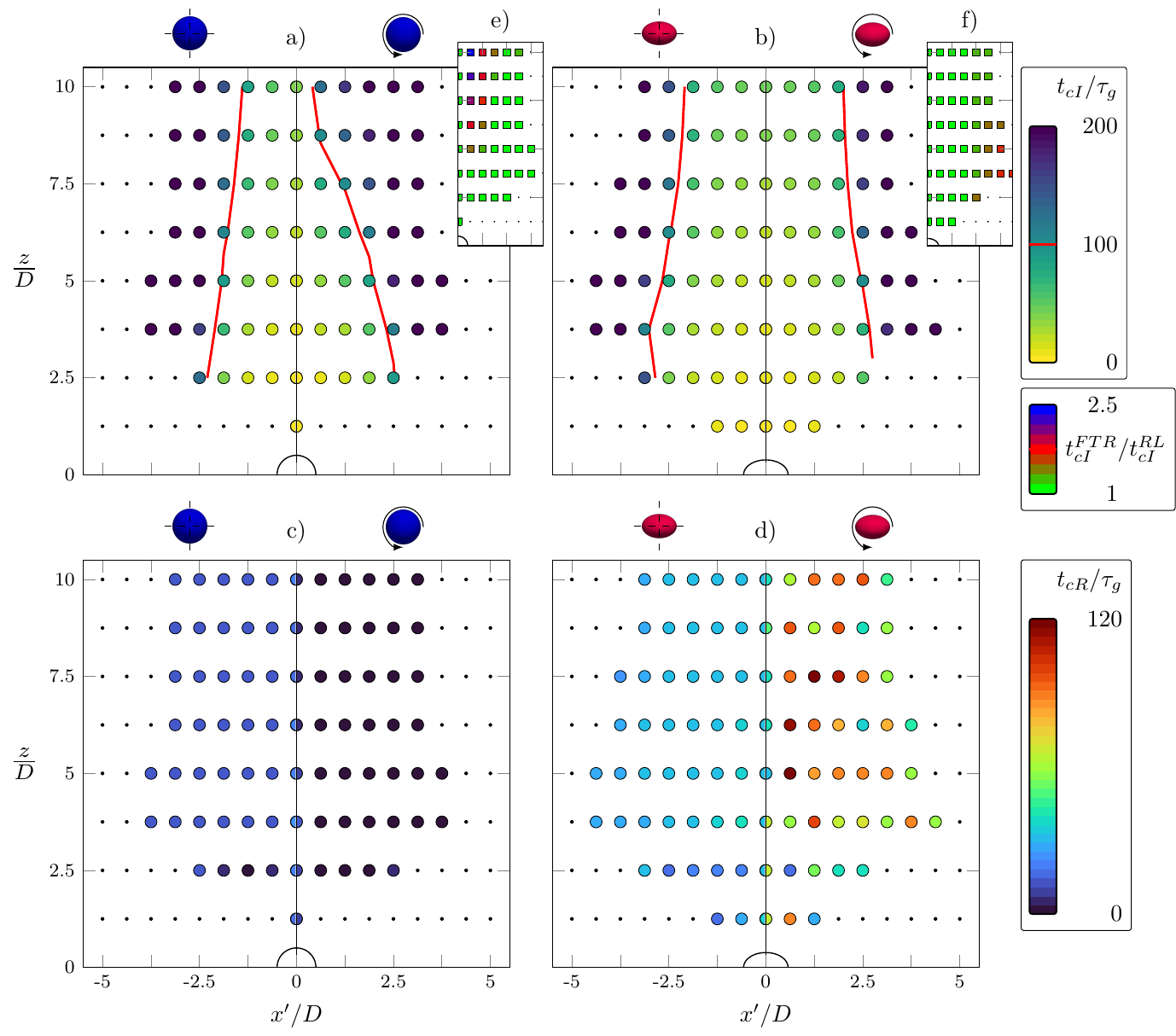} 
}
\caption{Maps of a,b) time to first collision \gls{mp:tkiss} and
c,d) interaction time of the \gls{dkt} cases as a function of the initial
condition of the trailing particle. The $x'$ axis for the rotationally-locked
cases is flipped to facilitate the comparison.
Cases in which no interaction occurred in the evaluated time are represented
with black dots and interacting cases are represented with coloured markers.
The red line in panels a,b is an isocontour of $\gls{mp:tkiss}=100\gls{tg}$.
The panels e,f contain the ratio of \gls{mp:tkiss} of free-to-rotate cases
with respect to their rotationally locked counterparts
($\gls{mp:tkiss;FTR}/\gls{mp:tkiss;RL}$).
\label{fig:DKT_maps}} 
\end{figure} 
Next we consider on the complete set of initial conditions of all four
configurations.
Figure \ref{fig:DKT_maps} shows maps of the time to first collision,
\gls{mp:tkiss}, and the interaction time, \gls{mp:tinter}, for the cases showing
at least one \gls{dkt} event.
The interaction time, \gls{mp:tinter}, measures the
time between the first collision and the smallest time after which the distance between
the particles grows monotonically (see precise definitions in \S~\ref{sec:dkt}),
evaluated as a function of the initial position of
the trailing particle.
The curve which delimits those initial conditions that lead to particle contact
from those which do not is very similar in all four cases (figure
\ref{fig:DKT_maps}a,b).
This means that the region of attraction in the explored range of relative positions is
essentially insensitive to the precise particle shape (sphere vs. mildly oblate
spheroid) and to their ability to rotate.
Both spheres and the spheroids considered, independently of angular motion being
suppressed or not, present a roughly cylindrical region of radius approximately
$3\gls{p:deq}$ located downstream of the leading particle in which particles will
interact.
Now let us consider the time to first collision, \gls{mp:tkiss}, which can give
us insight into the particles' response to wake attraction mechanisms in an 
integral sense, as a function of the initial relative position of the particles
(figure \ref{fig:DKT_maps} a,b).
We find similar values of \gls{mp:tkiss} for free-to-rotate spheroids and
rotationally-locked spheroids and spheres, whereas free-to-rotate spheres
present somewhat larger values for the same initial conditions.
This result suggests that the angular motion of spheres causes these particle
pairs to approach more slowly than corresponding non-rotating spheres and oblate
spheroids.
The latter do not rotate continuously, instead the angular motion during
the approach phase is characterized by small
oscillations around the equilibrium position (with the symmetry axis 
vertically aligned).
For the sake of clarity we include the ratio of \gls{mp:tkiss} of free-to-rotate
particles with respect to their rotationally locked counterparts in
the auxiliary panels (e,f) for each initial condition.
It can be seen that free-to-rotate spheres with a small horizontal shift in 
their relative position show ratios as large as $2.5$.
Furthermore, this ratio increases with the vertical distance within the range of
parameters evaluated.
These differences disappear for the cases
whose initial relative position has almost no horizontal shift, indicating that a
larger horizontal shift is required to trigger the rotation of the trailing sphere.
Next let us focus on the interaction of particle pairs after the first
collision.
Only a maximum of one collision is observed in the entire data-sets for rotationally locked
spheres and spheroids and free-to-rotate spheres.
Free-to-rotate spheroids, however, present two or three collision events per 
parametric point for most of the chosen initial conditions, except for a few 
of them in which a single, or even four collisions occur.
Figure \ref{fig:DKT_maps} c,d shows the interaction time \gls{mp:tinter}, where
a clear ascending order of configurations with respect to the duration of
particle-pair interactions is identified:
i) free-to-rotate spheres present an almost negligible interaction time,
ii) rotationally locked spheres interact during approximately $5\gls{tg}$, the value
of \gls{mp:tinter} increases to approximately $20-40\gls{tg}$ for the
iii) rotationally locked spheroids, and it reaches values above of up to
$100\gls{tg}$ for the iv) free-to-rotate spheroids.
In the following we will analyze the temporal evolution of the distance
separating the two interacting particles after the first contact.
For this purpose we set the time of the first contact arbitrarily to zero, and
then average the data over the ensemble of realizations with different initial
separations.
We define the average post-collisional distance as follows:
\begin{equation}\label{eq:lavg}
L\left(\tilde{t}\right)=\frac{\sum_i^{N} %
\lVert \vec{x}^i_{r,trail}\left( \tilde{t}\right)\rVert}{N} \,,
\end{equation}
where $\vec{x}^i_{r,trail}$ represents, for each case $i$, the relative position
of the trailing particle with respect to the leading particle and $\tilde{t}$
measures the time after the first collision (recall from figure \ref{fig:DKT_maps}$(c,d)$
that the value of \gls{mp:tkiss} is case-dependent).
Figure \ref{fig:DKT_avgdistance_all} shows the average post-collisional
distance, $L\left(\tilde{t}\right)$, for the four investigated configurations.
First, let us compare rotationally locked spheres versus rotationally locked 
spheroids.
On average, rotationally locked spheres drift away from each other a distance
of approximately $2.5\gls{p:deq}$ in the first $10\gls{tg}$ after the initial contact
(figure \ref{fig:DKT_avgdistance_all}b).
The scenario for rotationally locked spheroids is, on average, slightly more
complex.
There is a local maximum of the inter-particle distance at 
$\tilde{t}\approx 20\gls{tg}$, after which particles start to approximate each
other again, and a local minimum of $L$ at $\tilde{t}\approx 30\gls{tg}$, after which
particles drift away from each other.
Second, let us focus on the effect that angular motion has on both particle
shapes considered.
For the spheres, the angular motion results in particles staying much 
further away from each other after the collision, with an approximately constant
offset of around $1\gls{p:deq}$ for $\tilde{t} > 10\gls{tg}$.
This considerable effect due to the spheres' rotation can be understood
from figure \ref{fig:DKT_animated} which depicts the relative motion of particle
pairs for exemplary cases, as will be discussed in more detail below.
Conversely, when angular motion is allowed for spheroids, 
particles remain on average closer to each other at times larger 
than approximately $10\gls{tg}$ after the first collision, and 
the local maximum of $L$ observed for the rotationally locked counterparts is
more pronounced.
\begin{figure} 
\begin{center}
\includegraphics[scale=1.0]{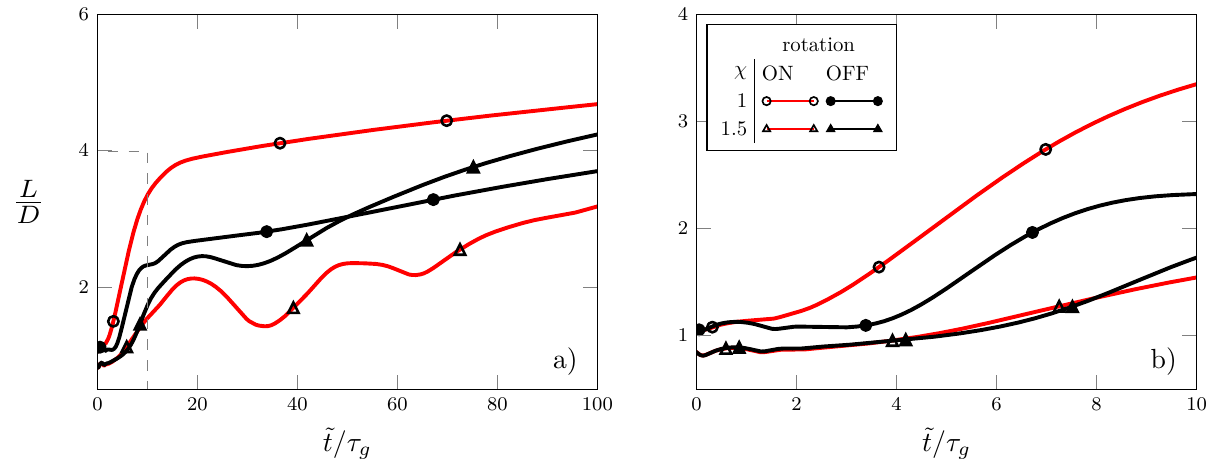} 
\end{center}
\caption{a) Time history of average distance \eqref{eq:lavg} 
after the first contact for the four configurations considered in the
\gls{dkt} configuration (see legend).
Non-colliding cases are excluded from the plot.
b) Zoom of panel a (see dashed rectangle in a).
\label{fig:DKT_avgdistance_all}} 
\end{figure}

\begin{figure} 
\makebox[\textwidth][c]{ 
\includegraphics[scale=0.8]{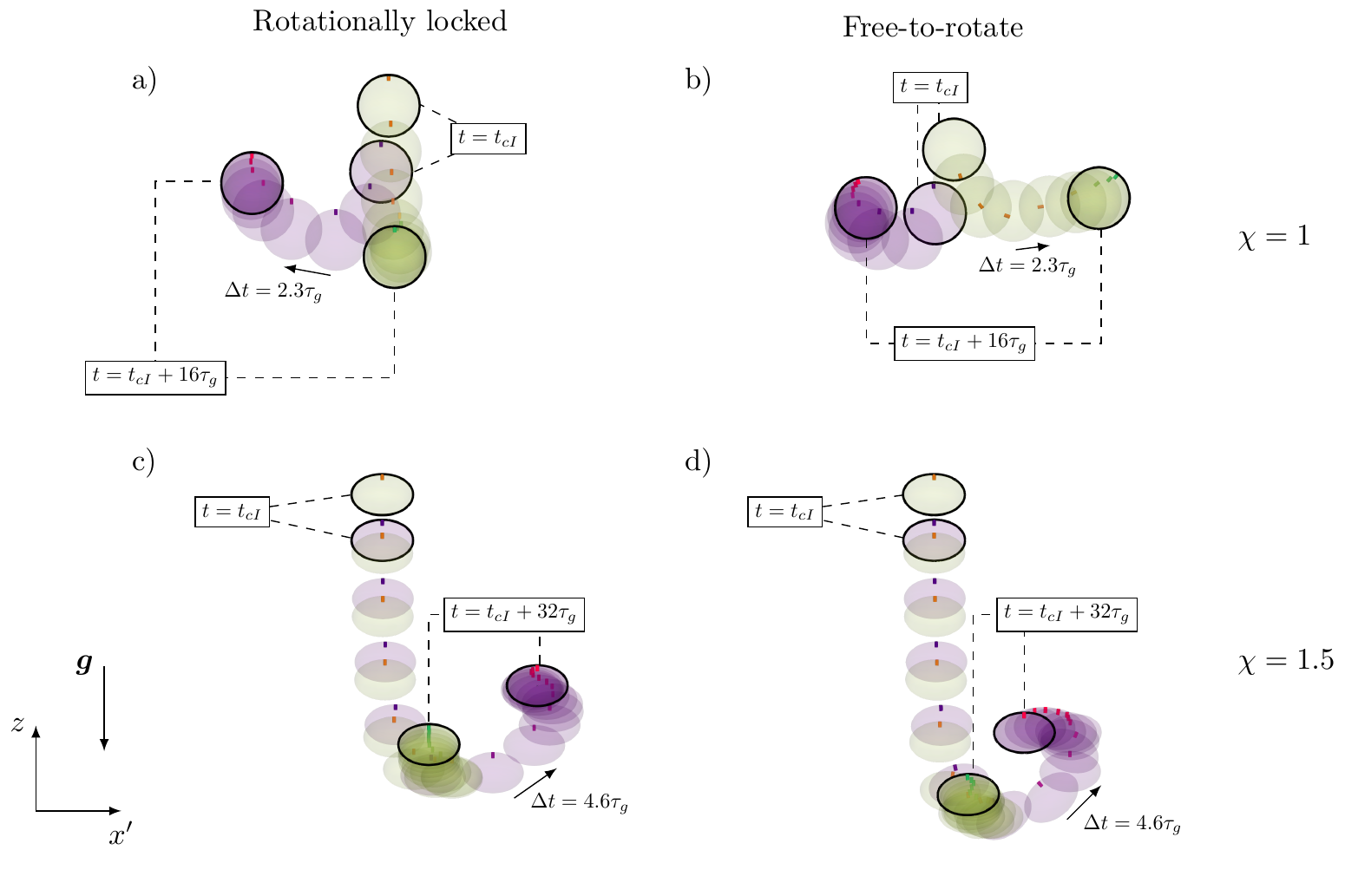} 
}
\caption{Overlay of consecutive snapshots of the different \gls{dkt} configurations for the
cases whose trailing particle starts at $\vec{x}_r=(0.625,7.5)\gls{p:deq}$.
The initial and final snapshots are indicated by highlighting the particles' contour.
The time interval selected is such that starts at the first contact
($t=\gls{mp:tkiss}$) and ends after $16\gls{tg}$ for spheres and $32\gls{tg}$
for spheroids, sampling $8$ equispaced time instants.
The time between consecutive snapshots $\Delta t$ is indicated in the figure.
The reference frame is translating downwards at a speed slightly smaller than
the settling velocity of a single particle ($0.975\gls{wref}$).
Particles are identified by color (trailing: green, leading: purple).
Time and angular position are indicated with a small mark whose color changes
with time.
\label{fig:DKT_animated}} 
\end{figure} 

Let us now identify possible phenomena that lead to the strong differences
in the pairwise interaction times after the first collision.
First, the larger separation observed in spheres as compared to spheroids just
after the first collision may be attributed to the mechanism by which pairs of
particles start to tumble after they contact each other.
According to \cite{fortes:1987}, vertically aligned spheres which are in contact
form an unstable system which is the actual cause of the tumbling phase.
This instability is due to the fact that two spheres vertically aligned can be
seen as an elongated body.
Such body would tilt so that it settles maximizing drag.
Since the pairs of particles considered are not connected, they separate from
each other shortly after they start to tilt.
The ratio between height and width of the body formed by two vertically aligned
and touching spheres is equal to 2, while the same quantity for spheroids (with
their axis of symmetry aligned with the vertical) equals $2/\chi$, which amounts
to $4/3$ in the present case.
Therefore, we can expect a more abrupt initial tumbling motion in the case of 
spheres.
Please note that if the oblate spheroids considered had an aspect ratio of 
$\gls{chi}=3$, as considered by \cite{fornari:2018b}, the ratio between height
and width becomes $2/3$, which is smaller than unity. 
In such a situation the tandem of vertically aligned particles would be even
more stable. 
Indeed, \cite{fornari:2018b} reported a stick mechanism in which the oblate spheroids
stay together, suppressing the tumbling phase.
Regarding the angular motion, two different effects are considered as candidates
to explain the lower values of the particle-pair separation distance $L$ for
spheroids and higher values of $L$ for spheres, when compared to their
rotationally-locked counterparts.
In the case of spheroids the angular motion allows the particles to have a
stronger rocking motion around a horizontal axis after their first collision.
This rocking motion is a consequence of the leading particle tilting when pushed
downwards by the trailing particle.
Figure \ref{fig:DKT_animated}d shows the tilting of the leading particle
and the consequent rocking motion after the first collision. 
This results in a zig-zag trajectory and, as a consequence, particles
stay, on average, closer to each other, thereby increasing the probability for
repeated collisions.
The subsequent collisions show similar features, but the amplitude of their
rocking motion is decreased, and the distance between the particles after 
the collisions is increased.
In the case of spheres, the distance between centers of free-to-rotate spheres
is approximately $1.5$ times larger than that of the rotationally-locked counterparts.
We attribute this to a rotation-induced lift force on the trailing particle
that increases the distance between the two particles.
This rotation can be appreciated in figure \ref{fig:DKT_animated}b, and its
effect on the relative trajectory between the particles is visible in figure 
\ref{fig:DKT_trajectories_rel}a.

\section{Discussion}
\label{sec:discussion}

\begin{figure} 
\begin{center}
\includegraphics[scale=0.9]{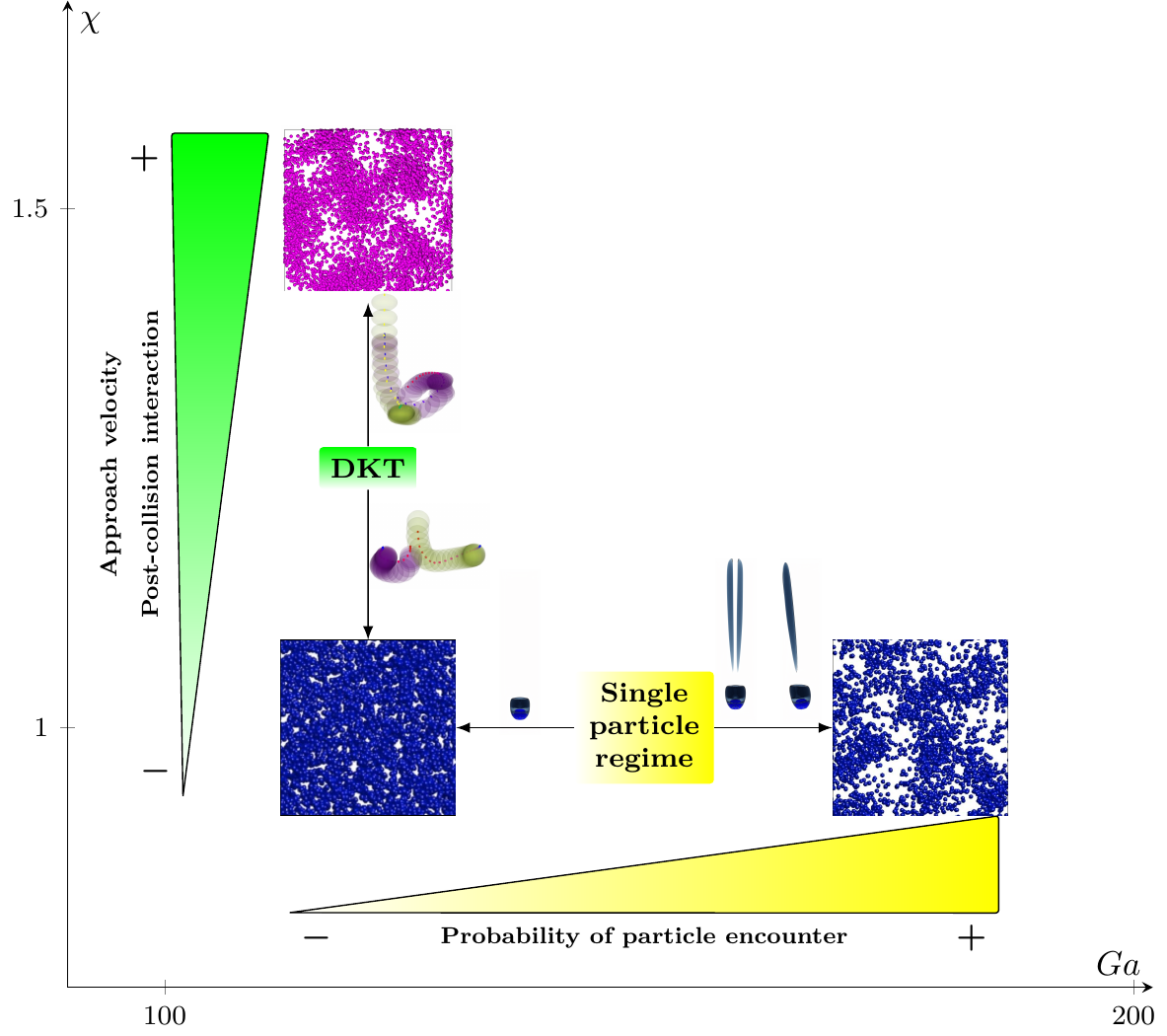} 
\end{center}
\caption{Summary sketch of clustering mechanisms analyzed in this work (intense
\gls{dkt} interactions) and from the reference work of \cite{uhlmann:2014a} 
(promoted particle encounters by horizontal motion).
\label{fig:clustering_scheme}} 
\end{figure} 

On the one hand our present many-particle simulations show that mildly
oblate spheroids (with an aspect ratio of $1.5$) form strong clusters
at a comparably low value of the Galileo number, for which an isolated
particle in ambient fluid settles in the steady vertical regime. 
This observation is in contrast to the known behavior of ensembles of
spheres, for which the Galileo number has been identified as a
critical parameter with respect to the onset of wake-induced
clustering \citep{uhlmann:2014a}. This clustering transition in the case
of spheres occurs in the range $\gls{Ga}=121\ldots178$ which encloses the
value $\gls{Ga}=155.8$ at which an isolated particle's path regime
bifurcates from steady vertical to steady oblique \citep{zhou:2015}.
The explanation proposed by \cite{uhlmann:2014a} links the onset of
clustering in the case of spheres to the enhanced horizontal mobility
of individual particles above the oblique-path threshold, which causes
an increase of the frequency of particle-particle encounters.
In order to make this point clearer, let us consider the fact that in
the wake of a given particle a ``region of attraction'' can be
defined, i.e.\ a spatial region of limited extent within which a 
trailing particle near equilibrium will experience a modified
hydrodynamic force which has the effect that -- without further
perturbations -- it will continuously approach the leading particle
until contact.
Now let us consider the hypothetical case of a set of mono-disperse
particles which are initially all located outside of any other
particles' region of attraction. If these particles are settling in
the steady vertical regime in an unbounded ambient fluid, they simply
move as a group without changing their relative positions, and they 
will, therefore, never get into contact. 
If, however, each individual particle of this ensemble settles in a
steady oblique regime (with a random azimuthal angle), there exists a
finite probability for individual particles to encounter another
particle's region of attraction, and to get into direct contact. If
the conditions are such that an interacting particle pair (on average) 
encounters one (or more) additional particle(s) before separating
again, initial seeds will eventually lead to large-scale cluster
formation through accretion. 
Hence, we can conclude that horizontal particle mobility can have the
effect of enhancing the tendency to cluster in this scenario. 
However, as we have observed in the present many-particle simulations,
oblate spheroids with aspect ratio $\gls{chi}=1.5$ apparently do not
require this mechanism in order to form columnar clusters.
This is in accordance with the results of 
\cite{fornari:2018b} for oblate spheroids at larger aspect ratio
($\gls{chi}=3$), lower Galileo number ($\gls{Ga}=60$) and smaller density ratio
($\gls{kappa}=1.02$) which likewise appear to form clusters. 

On the other hand, we have seen in the present series of
drafting-kissing-tumbling simulations that pairs of mildly
oblate spheroids, which are initially positioned such that the
trailing particle is located inside of the leading particle's region
of attraction, interact for a significantly longer time than
spherical counterparts at corresponding parameter points.
We have observed that the difference in interaction time is caused by
two factors: first, spheroids approach each other at a somewhat faster
rate than spheres during the drafting phase; second, spheroids
remain in close proximity over a significantly longer duration after
the first contact.
This is qualitatively in line with the results of
\cite{ardekani:2016} in their DKT simulation of spheroids with larger
aspect ratio ($\gls{chi}=3$), smaller Galileo number ($\gls{Ga}=80$) and smaller
density ratio ($\gls{kappa}=1.14$). These authors observe that those
flatter objects do not tumble, but instead remain locked in position
as a stack after the initial contact. 
Translating our result on the isolated pairwise interaction to the
(dilute) many-particle configuration implies that the observed
increase in the interaction time between pairs of particles can also
enhance the tendency to form large-scale clusters. This is due to the
fact that once a pair of particles gets into contact, it will have a
higher probability of attracting an additional particle before
separating, if the original pair has an extended contact duration. 
Hence, this argument suggests that an increase in the interaction time
between pairs of particles can lead to large-scale clustering, which is a 
mechanism that is different from (and complementary to) the above
process by enhancement of lateral mobility. 

Figure~19 shows a sketch which is intended to contrast the above two
routes of cluster formation in the dilute regime for
sphere-like particles. 
Starting from the (non-clustering) baseline case with spherical
particles at a Galileo number of ${\cal O}(100)$ in the steady
vertical regime (lower left corner of the $(\gls{Ga},\chi)$-plane in
figure~19), clustering can be enabled by either one of the following
two options:
(i) by increasing the probability of particles entering their
peers' attractive wake region (which is achieved by increasing the
horizontal particle mobility, i.e.\ through increasing the spheres'
Galileo number \gls{Ga}); 
(ii) by increasing the temporal interval over which particles remain
close to each other during wake-induced encounters (which can be
achieved by replacing spheres with oblate spheroids, i.e.\ by way of
increasing the particles' aspect ratio \gls{chi}).
Please note that additional triggers of cluster formation (besides
lateral particle mobility and close interaction duration) might be
at play, such as the growth of the spatial extent of the wake with
increasing Galileo number.
In addition, it remains to be understood how the density ratio influences
the mechanisms mentioned above.
Further research would be necessary to properly answer these questions.

\section{Conclusions}
\label{sec:conclusions}
We have performed particle-resolved direct numerical simulations
(PR-DNS) of many heavy non-spherical particles settling under gravity.
The particles are oblate spheroids of aspect ratio $1.5$ (which represent a
modest deviation from a spherical shape) and density ratio
$\gls{kappa}=1.5$; the global solid volume fraction measures $0.005$
such that the suspension can be considered as dilute. 
Two Galileo numbers are considered, namely $111$ and $152$ for which a
single oblate spheroid
follows a steady vertical and a steady oblique path, respectively.
In contrast to previous results for spheres \citep{uhlmann:2014a} we have found
that the qualitative difference in the single particle regime does not result in
a qualitatively different behavior of the multiparticle cases:
at both Galileo number values a strongly inhomogeneous spatial
distribution of the disperse phase in the form of columnar clusters is
observed, with a significantly enhanced average settling velocity as a
consequence.
A similar result has previously been reported for PR-DNS of
significantly flatter spheroids (with $\gls{chi}=3$) and for lower
density ratio ($\gls{kappa}=1.02$) by \cite{fornari:2018b}.  
Here we have used Voronoi tessellation as a basis for the
analysis of the structure of the particulate phase. The intensity of
clustering has been measured with the aid of the standard deviation of
the Voronoi cells' volume, normalized with the value obtained from a
random Poisson process \citep{monchaux:2010b}. It turns out that the
enhancement of the average settling speed is approximately
proportional to the standard deviation of the Voronoi cell volumes
when considering both the present spheroids as well as the spheres of
\cite{uhlmann:2014a} as a joint data-set. This result, however,
requires further confirmation through additional data points before
its potential implications can be evaluated. 
Note that the amount of enhancement of the settling velocity may still
depend on the domain size, since the size of particle clusters approaches
the former.

Motivated by the 
lack of influence of
the single-particle regime
upon
the statistical features of the multi-particle settling 
we have carried out a thorough analysis of
pairwise interactions of particles in the well-known drafting-kissing-tumbling
setup,
conducted in a computational domain with inflow/outflow boundary
conditions in the vertical direction. 
We have considered four configurations, namely oblate spheroids of aspect ratio
$1.5$ and spheres, with and without suppression of the angular motion, with density
ratio $\gls{kappa}=1.5$ and a Galileo number such that a single particle would 
follow a steady vertical path ($\gls{Ga}=111$).
.
Through systematic variation of the particle pair's relative initial position 
we have found that the region of attraction for both particle shapes, with and 
without rotation, is very similar. However, in the case of
free-to-rotate spheres the trailing particle's trajectory is
horizontally shifted towards larger radial distances, resulting 
in a prolonged drafting phase.
Regarding the particles' tumbling phase, we have shown that spheres
and spheroids behave in a qualitatively different manner.  
Spheres undergo at most a single collision, and they quickly separate
afterwards. 
Rotationally locked spheroids also experience a maximum of one
collision, but they remain close to each other for relatively long times.
Finally, free-to-rotate spheroids exhibit two collision events in most
of the cases in which a \gls{dkt} event is observed, and,
consequently, their average interaction time is the maximum out of the
four investigated configurations. 

To summarise, we observe a shape-induced
increase in the interaction time when two particles happen to ``meet''
(i.e.\ when one of them enters the other particle's wake region), such
that the probability of additional particles joining the initial pair
is increased with respect to the baseline case of spheres. 
Hence, the tendency to form a large-scale cluster increases. 
This is in contrast
to the mechanism for spheres \citep[as proposed by][]{uhlmann:2014a}, 
where the clustering transition is believed to be triggered by the
primary bifurcation of the isolated particle's wake flow (from
axisymmetric vertical to planar oblique) that leads to lateral
mobility, hence increasing the probability of mutual particle
encounters. 
As a consequence of these two observations, we conclude  
that the mechanism for the initiation of columnar clusters in the case of a dilute
suspension of modestly oblate spheroids is in a sense orthogonal to
the mechanism that is believed to be at work in the counterpart with
spherical particles.

Since the qualitatively different behavior of spheroids (as compared
to spheres) has now been established both for moderately flat
geometries \citep[$\gls{chi}=3$, ][]{fornari:2018b} and for modestly
flat ones ($\gls{chi}=1.5$, present work), the question of a possibly 
finite critical aspect ratio poses itself naturally. Future work
should be aimed in this direction.
Furthermore, a thorough analysis of collective effects in prolate
spheroids in a comparable parameter range is still lacking.

\appendix

\section{Collision model}
\label{sec:collision}

Here we describe the algorithm used to handle particle-particle contact in this
work. 
For the present case of a dilute suspension we adopt a simple repulsion model in
which contact forces are determined from the distance separating a given
particle pair.
Each contact event involves only a pair of particles, and the resultant contact force 
is assumed to be a point force.
Hence we need to define a contact point, a direction and a force intensity.
In this work we consider only normal forces with a quadratic law similar to the
one used in \cite{uhlmann:2014a} for spherical particles, originally proposed by
\cite{glowinski:2001}.
The main issue when working with non-spherical particles is that the contact
point and the normal direction are not uniquely defined.
The most popular methods to determine the contact parameters between spheroids
in the literature are the common normal \cite{lin:1995} and the geometric 
potential \citep{ng:1994}.
The common normal method is very attractive since it naturally yields the contact
point and the normal direction.
However, the resultant system of equations is under-determined and undesired solutions
can be obtained.
\cite{kildashti:2018} overcame this issue by an iterative process.
There is, however, a non-solved issue which arises when the overlapping distance
between the spheroids is exactly zero, or very small, leading to an
ill-determined system.
On the other hand, the geometric potential approach is particularly attractive
when dealing with simple geometries like spheroids, in which the contact point 
is easily determined.
The main drawback of the geometric potential is the definition of the normal
direction.
In this work we propose to use the geometric potential to determine the contact
point, but determine the normal direction with a slight modification of the 
algorithm originally proposed by \cite{ng:1994}.

\begin{figure} 
\makebox[\textwidth][c]{ 
\includegraphics[scale=1.3]{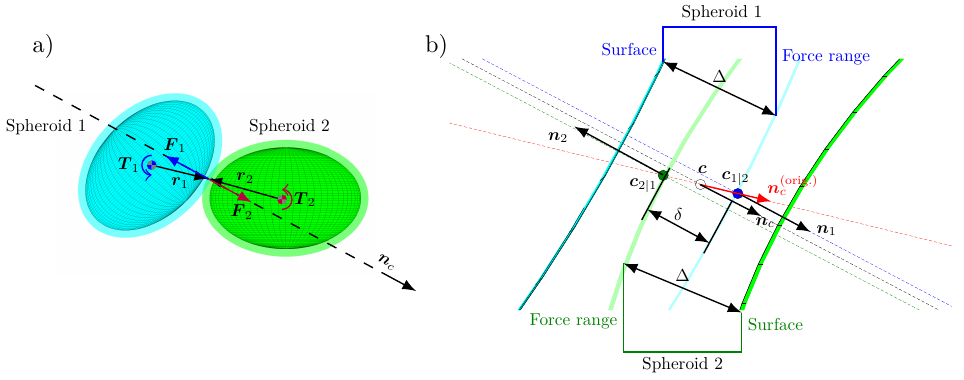} 
}
\caption{a) Sketch of the two spheroids indicating the contact force and torque
in each particle and the normal direction at the contact point.
b) Sketch of the elements involved in determining the contact point
(\gls{c:gp_cp}), the normal direction at the contact point (\gls{c:nn}) and
the overlapping distance ($\delta$).
The normal direction given by the original method  \citep[][%
$\vec{n}_c^{(\text{orig.})}$ in panel b %
]{ng:1994} is also including for comparison purposes.
\label{fig:contact_detail}} 
\end{figure}

In order to apply the geometric potential method we consider the \gls{cfr} of a
spheroid using the potential \gls{c:pot} defined as
\begin{equation}\label{eq:cfr}
\gls{c:pot}(x,y,z) = \gls{c:cfrA}x^2 + \gls{c:cfrB}y^2 + \gls{c:cfrC}z^2 
       +2\gls{c:cfrF}yz  +2\gls{c:cfrG}zx  +2\gls{c:cfrH}xy 
       +2\gls{c:cfrP}x   +2\gls{c:cfrQ}y   +2\gls{c:cfrR}z  + \gls{c:cfrD} \,,
\end{equation}
where the coefficients $\{\gls{c:cfrA}, \gls{c:cfrB}, \gls{c:cfrC}, \gls{c:cfrD},
\gls{c:cfrF}, \gls{c:cfrG}, \gls{c:cfrH}, \gls{c:cfrP}, \gls{c:cfrQ}, \gls{c:cfrR} \}$
are functions of the spheroid parameters (equatorial diameter $\gls{dd}$ and 
symmetry axis length $\gls{aa}$) and the particle's position and orientation.
For a point on the surface of the spheroid $\gls{c:pot}=0$.
The coefficients in \eqref{eq:cfr} are easily obtained from the \gls{cfr} of
the spheroid expressed in the body-fixed reference system
\begin{equation}\label{eq:cfr_body}
\gls{c:pot}(\gls{Obody_x},\gls{Obody_y},\gls{Obody_z}) = 
       \left(\frac{\gls{Obody_x}}{\gls{dd}/2+\gls{dx}/2}\right)^2 
     + \left(\frac{\gls{Obody_y}}{\gls{dd}/2+\gls{dx}/2}\right)^2 
     + \left(\frac{\gls{Obody_z}}{\gls{aa}/2+\gls{dx}/2}\right)^2 - 1  \,,
\end{equation}
where $(\gls{Obody_x},\gls{Obody_y},\gls{Obody_z})$ are the coordinates of a point 
\gls{Obody_vx} expressed in the body-fixed
coordinate system (see figure \ref{fig:problem_description}a) and where we have 
included a force range of $\gls{dx}/2$  in order to minimize the effect of overlapping
support of the diffuse interface during particle approach (see figure \ref{fig:contact_detail}b).
After some algebra  and using the relation 
$\gls{Obody_vx} = \gls{rotmat}\left(\gls{Oxyz_vx}-\gls{vxp}\right)$,
where \gls{rotmat} is the rotation matrix to obtain body-fixed coordinates from
the global coordinate system, one reaches to
\begin{subequations}
\begin{align}
\gls{c:cfrA} &= \sum_{i} U_i\gls{rotmat_i1}^2 \,,\\
\gls{c:cfrB} &= \sum_{i} U_i\gls{rotmat_i2}^2 \,,\\
\gls{c:cfrC} &= \sum_{i} U_i\gls{rotmat_i3}^2 \,,\\
\gls{c:cfrF} &= \frac{1}{2}\sum_{i} U_i\gls{rotmat_i2}\gls{rotmat_i3} \,,\\ 
\gls{c:cfrG} &= \frac{1}{2}\sum_{i} U_i\gls{rotmat_i3}\gls{rotmat_i1} \,,\\ 
\gls{c:cfrH} &= \frac{1}{2}\sum_{i} U_i\gls{rotmat_i1}\gls{rotmat_i2} \,,\\ 
\gls{c:cfrP} &=-\frac{1}{2}\sum_{r}\gls{vxp_r}\sum_{i} U_i\gls{rotmat_ir}\gls{rotmat_i1} \,,\\ 
\gls{c:cfrQ} &=-\frac{1}{2}\sum_{r}\gls{vxp_r}\sum_{i} U_i\gls{rotmat_ir}\gls{rotmat_i2} \,,\\ 
\gls{c:cfrR} &=-\frac{1}{2}\sum_{r}\gls{vxp_r}\sum_{i} U_i\gls{rotmat_ir}\gls{rotmat_i3} \,,\\ 
\gls{c:cfrD} &= \sum_{s}\gls{vxp_s}\sum_{r}\gls{vxp_r}\sum_{i} U_i\gls{rotmat_ir}\gls{rotmat_is} - 1 \,,
\end{align}
\end{subequations}
where $\vec{U} = \left(\left(\gls{dd}/2+\gls{dx}/2\right)^{-2},
                 \left(\gls{dd}/2+\gls{dx}/2\right)^{-2},
                 \left(\gls{aa}/2+\gls{dx}/2\right)^{-2}\right)$.
In the following we introduce the subscript $1$ or $2$ to identify each of the
spheroids participating in a collision, which are defined by their potentials
$\gls{c:pot}_1$ and $\gls{c:pot}_2$.
The contact point is defined as the midpoint between the deepest point of 
spheroid $1$ in $2$, \gls{c:gp_deep_1}, and the deepest point of spheroid $2$
in $1$, \gls{c:gp_deep_2}.
To obtain the deepest point of $1$ in $2$ we minimize the function $\mathcal{L}=
\gls{c:pot}_2 + \gls{c:gp_la}\gls{c:pot}_1$.
The following linear system is obtained
\begin{equation}\label{eq:lagrange}
   \left[
   \begin{array}{ccc}
   \gls{c:cfrA}_2 + \gls{c:gp_la}\gls{c:cfrA}_1 & \gls{c:cfrH}_2 + \gls{c:gp_la}\gls{c:cfrH}_1 & \gls{c:cfrG}_2 + \gls{c:gp_la}\gls{c:cfrG}_1 \\
   \gls{c:cfrH}_2 + \gls{c:gp_la}\gls{c:cfrH}_1 & \gls{c:cfrB}_2 + \gls{c:gp_la}\gls{c:cfrB}_1 & \gls{c:cfrF}_2 + \gls{c:gp_la}\gls{c:cfrF}_1 \\
   \gls{c:cfrG}_2 + \gls{c:gp_la}\gls{c:cfrG}_1 & \gls{c:cfrF}_2 + \gls{c:gp_la}\gls{c:cfrF}_1 & \gls{c:cfrC}_2 + \gls{c:gp_la}\gls{c:cfrC}_1
   \end{array}
	\right] \left[
		\begin{array}{c}
                       x_{\gls{c:gp_la}} \\
                       y_{\gls{c:gp_la}} \\
		       z_{\gls{c:gp_la}} 
		\end{array}
    	\right] = -\left[
		\begin{array}{c}
                        \gls{c:cfrP}_2 + \gls{c:gp_la}\gls{c:cfrP}_1 \\
                        \gls{c:cfrQ}_2 + \gls{c:gp_la}\gls{c:cfrQ}_1 \\
                        \gls{c:cfrR}_2 + \gls{c:gp_la}\gls{c:cfrR}_1 
		\end{array}
	\right].
\end{equation}%
We can obtain a solution for the linear system \eqref{eq:lagrange} in terms of 
\gls{c:gp_la} with Cramer's rule
\begin{equation}\label{eq:cramer}
x_{\gls{c:gp_la}} = \frac{\gls{c:gp_D1}}{\gls{c:gp_D}},  \quad
y_{\gls{c:gp_la}} = \frac{\gls{c:gp_D2}}{\gls{c:gp_D}},  \quad
z_{\gls{c:gp_la}} = \frac{\gls{c:gp_D3}}{\gls{c:gp_D}},  
\end{equation}
where \gls{c:gp_D}, \gls{c:gp_D1}, \gls{c:gp_D2} and \gls{c:gp_D3} are the determinants 
of the matrix of the linear system \eqref{eq:lagrange}.
Substituting \eqref{eq:cramer} in the potential of spheroid 1 leads to a sixth order 
polynomial for \gls{c:gp_la}
\begin{equation}\label{eq:sextic}
\gls{c:pot}_1\left(x_{\gls{c:gp_la}},y_{\gls{c:gp_la}},z_{\gls{c:gp_la}}\right) = 0.
\end{equation}
Now, we define \gls{c:gp_la_min} as the real-valued solution out of the set 
$\gls{c:gp_la}_i$ ($i=1,6$) for which $\gls{c:pot}(\gls{c:gp_la_min}) = 
\min_i \gls{c:pot}(\gls{c:gp_la}_i)$.
The deepest point of spheroid 1 inside spheroid 2 is $\gls{c:gp_deep_1}%
=(x_{\gls{c:gp_la_min}},y_{\gls{c:gp_la_min}},z_{\gls{c:gp_la_min}})$.
If $\gls{c:pot}_2(\gls{c:gp_deep_1}) > 0$, then contact does not exist, and the 
computation of \gls{c:gp_deep_2} is skipped.
If $\gls{c:pot}_2(\gls{c:gp_deep_1}) = 0$, $\gls{c:gp_deep_2}=
\gls{c:gp_deep_1}$.
If $\gls{c:pot}_2(\gls{c:gp_deep_1}) < 0$, the point \gls{c:gp_deep_2} is
obtained analogously to \gls{c:gp_deep_1}.
In the event of collision, the contact point \gls{c:gp_cp} is defined as 
the arithmetic average position
$\gls{c:gp_cp} = (\gls{c:gp_deep_1}+\gls{c:gp_deep_2})/2$.

The normal direction to the contact, \gls{c:nn}, according to the original
method is defined by the line connecting the points \gls{c:gp_deep_1} and 
\gls{c:gp_deep_2}, $\gls{c:nn} =\frac{\gls{c:gp_deep_2}-\gls{c:gp_deep_1}}%
{\left\|\gls{c:gp_deep_2}-\gls{c:gp_deep_1}\right\|}$.
However, the line connecting the two deepest points is not (in general) aligned
with the direction obtained from the common normal method.
Therefore, we propose to define the normal direction at the contact point as 
$\gls{c:nn} = (\gls{c:gp_nn_1} - \gls{c:gp_nn_2})/2$, where \gls{c:gp_nn_1} 
(\gls{c:gp_nn_2}) represents the unitary normal vector to spheroid 1 (2) at point
\gls{c:gp_deep_1} (\gls{c:gp_deep_2}).
It should be noted that \gls{c:nn} is not exactly normal to any of spheroids
1 or 2.

Finally, the modulus of the normal contact force is defined as
$F_n =\delta^2/J$, where $\delta$ is the overlapping distance ($\delta = %
\gls{c:nn} \cdot (\gls{c:gp_deep_1}-\gls{c:gp_deep_2})$) and $J$ is a constant
whose value depends on the submerged weight of a single particle, $W_s$, as
$J\,W_s/D^2\approx2.5\cdot 10^{-4}$
(we have checked that our DKT results are not sensitive to the precise choice of
the stiffness constant $J$).
This leads to the following forces and torques on each of the colliding two particles in a pair:
\begin{subequations}
\begin{align}
\gls{c:F1} &= -\gls{c:F2} = -F_n\gls{c:nn}\\
\gls{c:T1} &= \gls{rs:1} \times  \gls{c:F1}  \\
\gls{c:T2} &= \gls{rs:2} \times  \gls{c:F2}  \, .
\end{align}
\end{subequations}
Please note that, compared to spheres, normal forces can generate torque 
in the non-spherical case (see figure \ref{fig:contact_detail}a).
Furthermore, the pair of torques does not need be equal in magnitude.

\section{Two point autocorrelation functions}
\label{sec:autocorr}

\begin{figure} 
\makebox[\textwidth][c]{ 
\includegraphics[scale=1.0]{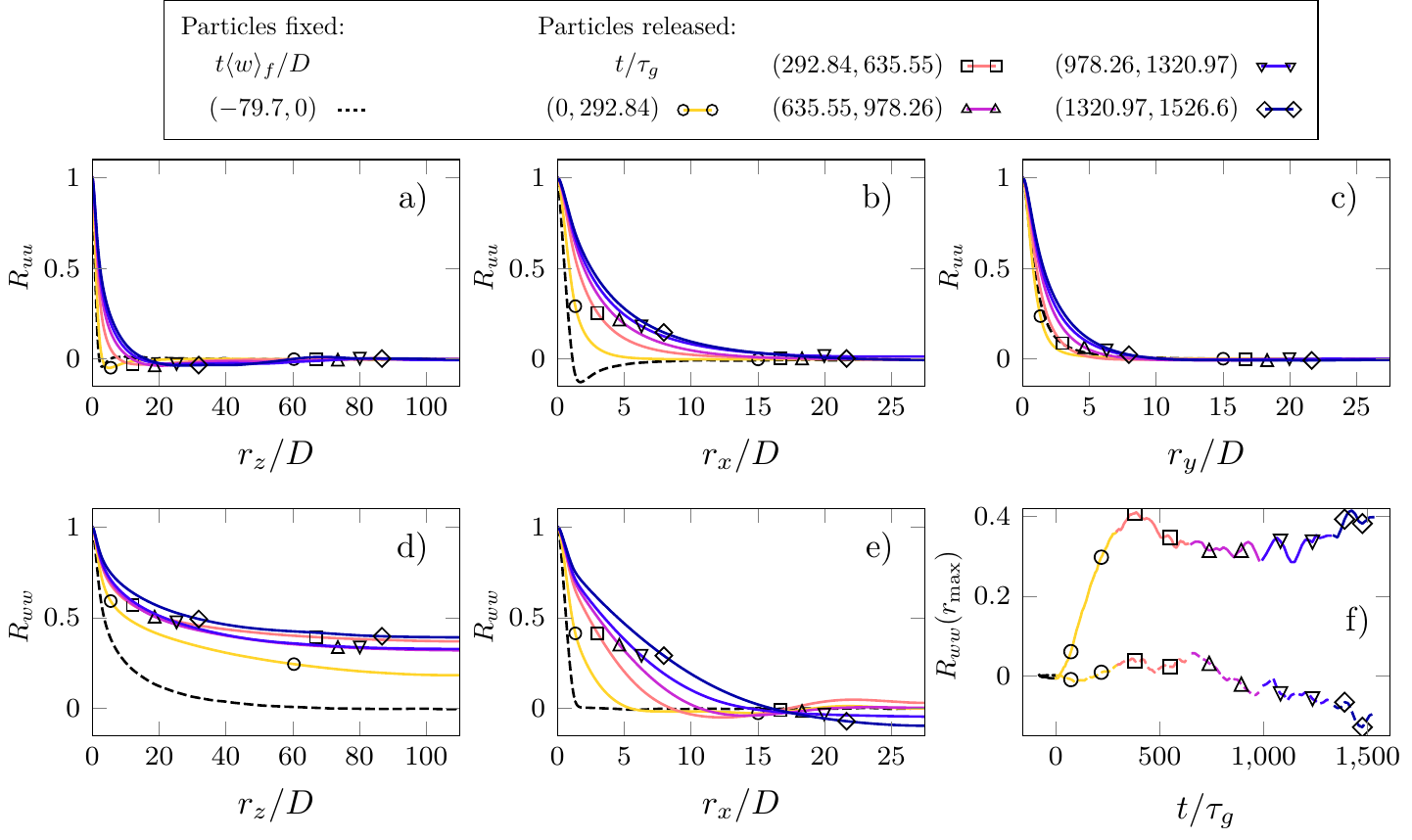} 
}
\caption{Autocorrelation functions of the
a-c) horizontal fluid velocity component , \gls{f:Ruu}, and of the
d-e) vertical counterpart, \gls{f:Rww},
for case {\tt G111}.
f) Time history of \gls{f:Rww} at the furthest vertical and horizontal position.
The time interval to compute each curve in panels a-e is shown in the legend,
and the corresponding linestyles are used piecewise in panel f.
\label{fig:autocorr_G111}} 
\end{figure}

\begin{figure} 
\makebox[\textwidth][c]{ 
\includegraphics[scale=1.0]{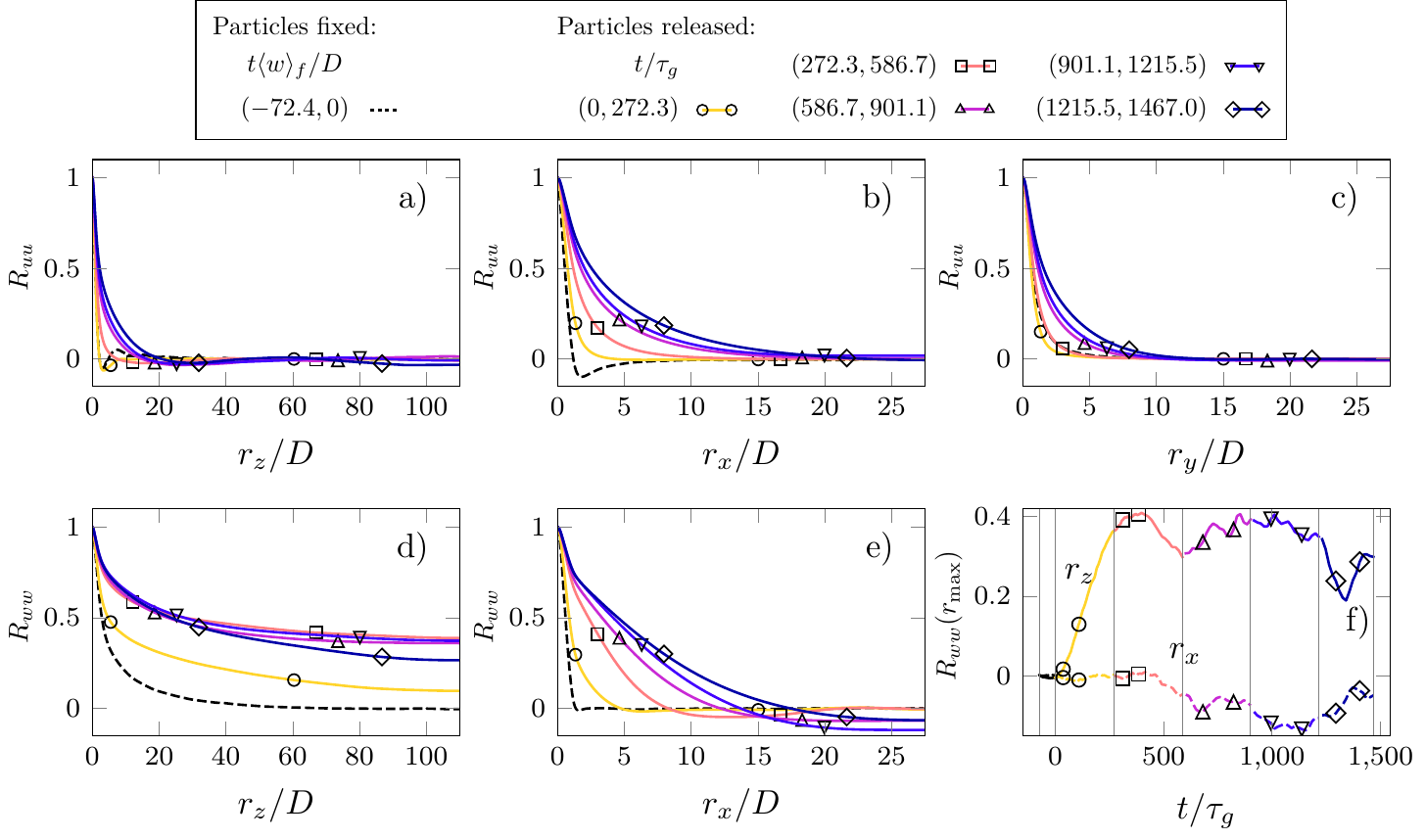} 
}
\caption{Same as \ref{fig:autocorr_G111} but for case {\tt G152}.
\label{fig:autocorr_G152}} 
\end{figure}

Figures \ref{fig:autocorr_G111} and \ref{fig:autocorr_G152} show the two point
autocorrelation functions 
of the \revision{vertical}{horizontal} fluid velocity component,
\revision{\gls{f:Rww}}{\gls{f:Ruu}}, (panels \revision{a and b}{a-c}) and of the
\revision{horizontal}{vertical} counterpart, \revision{\gls{f:Ruu}}{\gls{f:Rww}},
(panels \revision{d, e and f}{d-e}) for cases \verb!G111! and \verb!G152!, respectively, together with 
the time history of \gls{f:Rww} at the furthest vertical and horizontal 
position (panel \revision{c}{f}).
When particles are fixed ($t<0$) all signals are fully decorrelated within the
domain.
After particles are released the horizontal velocity remains fully decorrelated 
until the end of the simulated time (see figures \ref{fig:autocorr_G111}\revision{d-f}{a-c}
and \ref{fig:autocorr_G152}\revision{d-f}{a-c}), whereas the vertical velocity acquires
correlations which do not decay to zero at later times (figures 
\ref{fig:autocorr_G111}\revision{a-b}{d-e} and 
\ref{fig:autocorr_G152}\revision{a-b}{d-e}) as will be discussed in the following.
Along the vertical direction, \gls{f:Rww} presents an initial fast growth
($0\lesssim t/\gls{tg} \lesssim 300$).
For later times, the curves $\gls{f:Rww}(r_z)$ are similar, except for a small
oscillating behavior in time.
They show a monotonic decreasing behavior in space with positive values over the
entire domain ($0<r_z/D<110$).
This is the footprint of the fast formation of robust columnar structures that 
occupy the whole domain in the vertical direction.
The temporal evolution of \gls{f:Rww} at its furthest distance 
($r_{z,\text{max}}=110$) supports the fast growth and the small oscillation with
time commented above.

The behavior of \gls{f:Rww} along the horizontal direction is somehow more
complex than along the vertical direction.
First, it presents an almost decorrelated solution until $300\gls{tg}$ in case
\verb!G111! and $900\gls{tg}$ in case \verb!G152!.
In both cases there is a turnover point in which the value of \gls{f:Rww} 
and for case \verb!G111! even two when $300\lesssim t/\gls{tg}\lesssim 600$.
This can be explained by the growth of the clusters in the horizontal direction
being a slow process compared to their growth in the vertical direction.
Furthermore, once the clusters grow to a size comparable to the computational
domain and because of continuity, there are negative values at 
$r_{x,\text{max}}=55$.
The first crossing of \gls{f:Rww} with the zero value gives an estimate of the 
size of the clusters in the horizontal direction.
This measures approximately $15-20\gls{p:deq}$, which implies that the clusters
ultimately grow to a size for which the current computational domains are not sufficient.

\section{\glsentrylong{dkt} computational setup}
\label{sec:dkt}

Here we describe the computational setup of the \gls{dkt} simulations presented
in \S~\ref{sec:results/dkt}.
The problem description is the same as in the multiparticle cases 
(\S~\ref{sec:problem}), but the methodology presents a few differences compared
to the one presented in \S~\ref{sec:method}. 
First, we impose a free stream of constant velocity at the lower boundary plane,
an advective boundary condition at the top, and periodicity at the lateral
boundaries, instead of periodicity in the three spatial directions.
Second the number of particles is exactly two.
Thus, the solid volume fraction is not a parameter anymore, and the governing 
parameters of the \gls{dkt} cases are \gls{Ga} and \gls{kappa}.
We explore two different particle shapes, namely spheres and oblate spheroids
with $\gls{chi}=1.5$, both with $\gls{kappa}=1.5$ and $\gls{Ga}=110.56$.
The size of the computational domain measures $[10.66 \times 10.66 \times 21.33]%
\gls{p:deq}^3$, where \gls{p:deq} is the diameter of a sphere with the same volume
as the particle considered.
Finally, for each particle shape we perform simulations with angular motion
enabled or suppressed.

\begin{figure} 
\begin{center}
\includegraphics[scale=1.0]{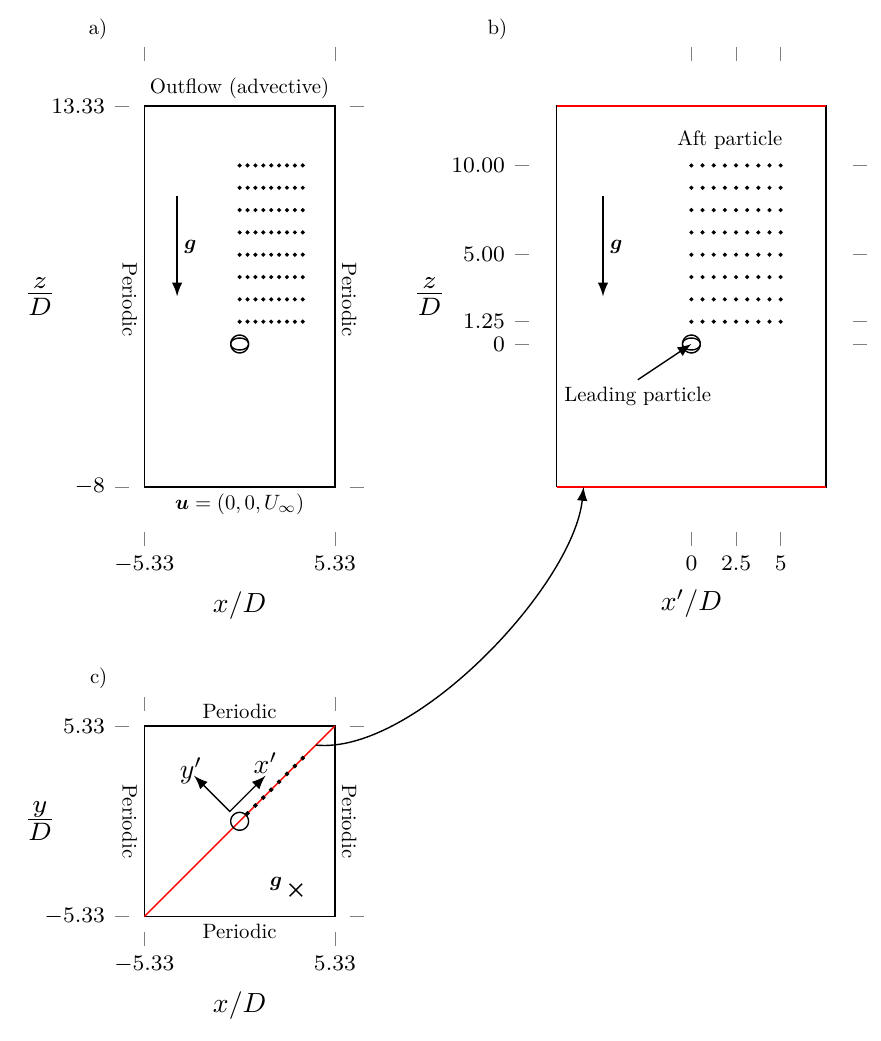} 
\end{center}
\caption{Outline of the \glsentrylong{dkt} simulations from a) lateral and c)
top views. b) View perpendicular to the plane where the trailing particle initial 
condition is located.
\label{fig:DKT_sketch}} 
\end{figure}

Figure \ref{fig:DKT_sketch} shows a sketch of the computational setup, indicating
the set of initial particle positions.
We refer to the particle which is initially at a lower vertical position as the 
leading particle, and to the other particle as the trailing particle.
The initial position of the leading particle is always at $8\gls{p:deq}$ above the 
bottom boundary of the computational domain.
The initial condition of the trailing particle is varied, sweeping an area of
$[5\times 8.75]\gls{p:deq}^2$ in the horizontal and vertical direction, 
respectively.
This area is uniformly sampled leaving an horizontal and vertical distance of 
$0.625 \gls{p:deq}$ and $1.25\gls{p:deq}$, respectively, between neighboring
initial conditions.
This results in $72$ simulations for each configuration, a total number of 
$288$ \gls{dkt} cases.
In order to reduce the influence of mirror particles due to periodicity, we 
locate the plane containing the initial condition of the trailing particle in the 
plane $x=y$.
We define the horizontal coordinate contained in this plane as $x'$ and 
the relative position of the trailing particle with respect to the leading particle 
in this plane as
\begin{equation}\label{eq:dkt/xr}
\vec{x}_r = \vec{x}_\text{trailing} - \vec{x}_\text{leading}.
\end{equation}

One of the key aspects in this computational setup (non-periodic in the 
direction of gravity) is to keep the particles inside the computational domain
for sufficiently long time intervals.
Therefore, we need a good estimate of the average settling velocity of the 
system (both particles), and a sufficiently large domain in the vertical 
direction to accommodate the variations of this settling velocity.
It should be mentioned that these variations can be large due to collision
events.
From trial simulations we found that imposing the reference Reynolds number
based on the free stream velocity imposed at the inlet $\gls{Reref}=\gls{Uref}%
\gls{p:deq}/\gls{nu}$ slightly higher than the terminal Reynolds number obtained
for a single particle in the same configuration leads to successful simulations.
From this experience we set $\gls{Reref}=1.025\gls{p:Redeq;0}$ in all the cases.
The following phases of the evolution of the system are identified:
\begin{enumerate}
   \item {\bf Initial upward drift}: if particles are initially at a sufficient
         distance away from each other, they slowly drift upwards in the computational
         domain because $\gls{Reref}>\gls{p:Redeq;0}$.
         For cases in which collisions do not occur, this is the only phase
         of the problem. For cases in which particles are close enough to each 
         other the \gls{dkt} event is triggered since the start of the simulations, and
         this phase is skipped.
   \item {\bf \gls{dkt} event}: if the trailing particle is attracted by the wake of
         the leading one, the former drafts towards the latter and they eventually 
         collide.
         This results in an enhancement of the settling velocity of both particles.
         Having drifted upwards in the previous phase leaves more clearance for the
         \gls{dkt} event to occur without encountering the bottom boundary.
   \item {\bf Final upward drift}: after the \gls{dkt} event both particles end
         up at a similar height and both drift upwards while repelling each
         other.
\end{enumerate}

We have verified that all the non-colliding cases have been run for at least
the maximum time to first interaction observed ($678\gls{tg}$, from rotationally-locked
spheroids with relative initial position of the trailing particle $(3.125,10)\gls{p:deq}$)
plus $100\gls{tg}$.
This results in all the cases simulated for at least $778\gls{tg}$.

\begin{figure} 
\makebox[\textwidth][c]{ 
\includegraphics[scale=1.0]{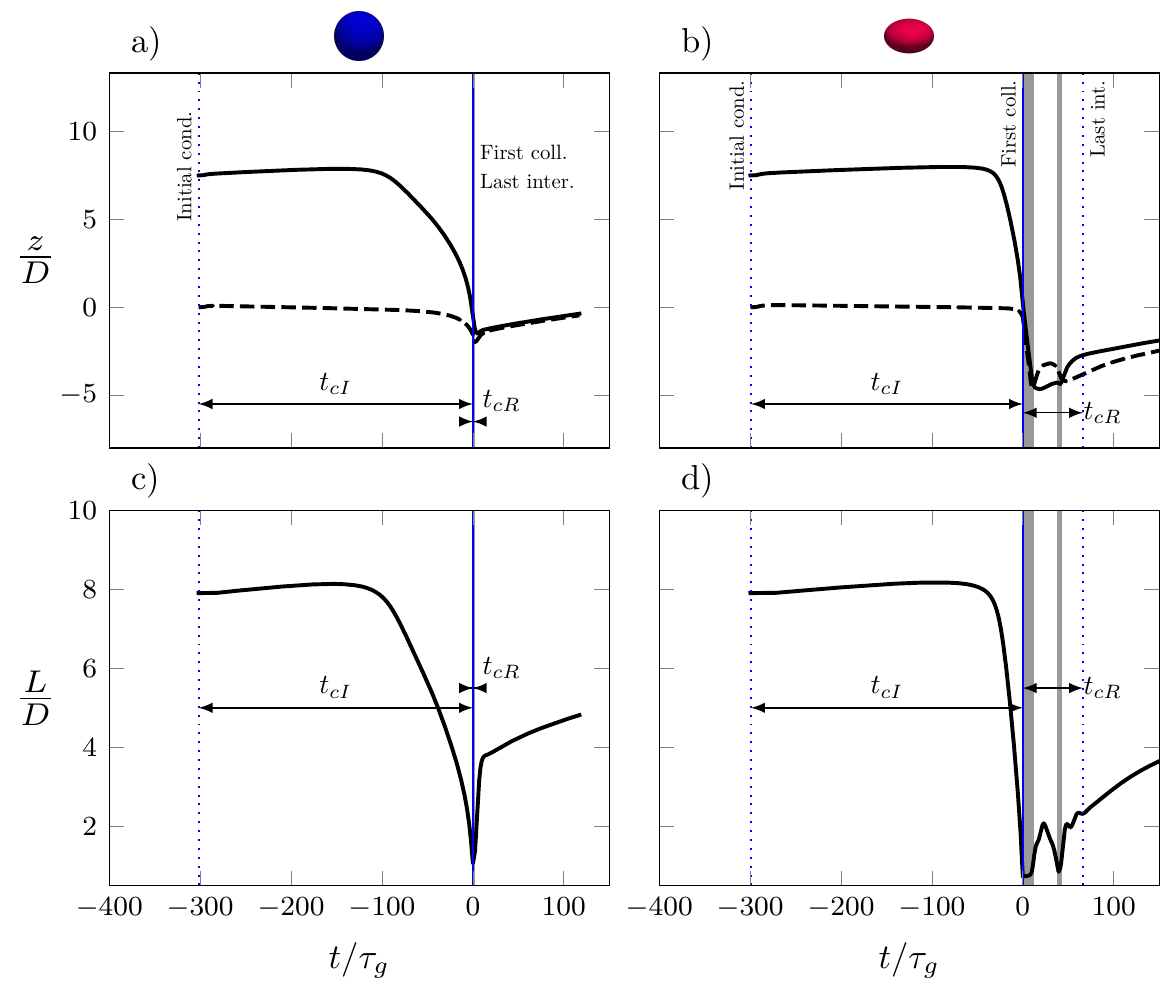} 
}
\caption{Time history of the vertical positions of a) spheres and b) spheroids
of $\gls{chi}=1.5$ and the distance between particle centers (c,d).
In both cases $(x_r/\gls{p:deq},y_r/\gls{p:deq})=(2.5,7.5)$ and the time is shifted so
that the instant of the first collision is $t=0$.
In (a-b) the trailing particle is represented with a solid line and the leading particle
with a dashed line. 
The gray shading indicates the time interval in which particles are in contact.
The vertical dotted lines illustrate the definition of the time to first collision,
\gls{mp:tkiss}, and the interaction time, \gls{mp:tinter}.
\label{fig:DKT_timehistory}} 
\end{figure}

Figure \ref{fig:DKT_timehistory} shows the time history of two \gls{dkt} 
simulations as an example. 
In both cases the angular motion is enabled and the initial condition of the 
trailing particle is $(x'_r/\gls{p:deq},z_r/\gls{p:deq})=(2.5,7.5)$.
For every simulation in which a collision takes place, we define the time 
to first collision, \gls{mp:tkiss}, and the interaction time \gls{mp:tinter}.
The former is the time between the begin of the simulation and the first
collision.
The latter is defined as the time between the first collision and the last 
time instant in which the particle centers approach each other.
The definition of these two quantities is indicated in figure 
\ref{fig:DKT_timehistory}.
Please note that for most of the cases of spheres with angular motion enabled
the interaction time is almost negligible (see figure 
\ref{fig:DKT_timehistory}c).

\section*{Supplementary data} 

\label{SupMat}Supplementary material (animations) have the digital object identifier
\href{https://dx.doi.org/10.5445/IR/1000151148}{\bf 10.5445/IR/1000151148}.

\section*{Acknowledgements}
The computations were partially performed on the
supercomputers ForHLR and HoreKa funded by the Ministry of Science, Research and the
Arts Baden--W\"urttemberg and by the Federal Ministry of Education and Research.

\section*{Funding}
This work was supported by the German Research Foundation (DFG)
under Project UH 242/11-1.

\section*{Declaration of interests}
The authors report no conflict of interest
\section*{Author ORCID}
M. Moriche,           \url{https://orcid.org/0000-0003-2855-8084}; 
M. Garc\'ia-Villalba, \url{https://orcid.org/0000-0002-6953-2270};
M. Uhlmann,           \url{https://orcid.org/0000-0001-7960-092X};

\bibliographystyle{authoryear} 
\addcontentsline{toc}{section}{References}
\bibliography{biblio}

\end{document}